\documentclass[12pt]{report}

\usepackage[latin1]{inputenc}
\usepackage[english]{babel}
\usepackage{graphicx}
\usepackage{makeidx}
\usepackage{indentfirst}
\usepackage{subeqn}

\title{}
\author{}
\date{}

\newcommand{\be}{\begin{equation}}
\newcommand{\ee}{\end{equation}}
\newcommand{\bea}{\begin{eqnarray}}
\newcommand{\eea}{\end{eqnarray}}
\newcommand{\bd}{\begin{displaymath}}
\newcommand{\ed}{\end{displaymath}}
\newcommand{\beastar}{\begin{eqnarray*}}
\newcommand{\eeastar}{\end{eqnarray*}}
\newcommand{\bmlett}{\begin{mathletters}}
\newcommand{\emlett}{\end{mathletters}}

\newcommand{\bit}{\begin{itemize}}
\newcommand{\eit}{\end{itemize}}
\newcommand{\ben}{\begin{enumerate}}
\newcommand{\een}{\end{enumerate}}
\newcommand{\bde}{\begin{description}}
\newcommand{\ede}{\end{description}}
\newcommand{\bce}{\begin{center}}
\newcommand{\ece}{\end{center}}

\newcommand{\sot}{\underline}

\newcommand{\lgr}{\left\{}

\newcommand{\lbl}{\left.}
\newcommand{\rbl}{\right.}
\newcommand{\lat}{\left|}
\newcommand{\rat}{\right|}
\newcommand{\fr}{\frac}

\newcommand{\tend}{\rightarrow}

\newcommand{\Tr}{\ensuremath{{\rm Tr}}}

\makeindex

\begin{document}

\pagestyle{empty}
\chapter*{Foreword}
\renewcommand{\thepage}{{}}

 This 1981 doctoral
thesis is posted on the ArXiv with  explicit written permission of
the Massachusetts Institute of Technology.

We post  it here because of  the following reasons:

\begin{itemize}
\item the Sc.D. thesis is referred to explicitly in a series of
articles (see below);
\item the initial chapters of the thesis contain still unpublished ideas and
results that are relevant to ongoing discussions on fundamental
problems in quantum theory, second-law violations, mixed versus
pure or heterogeneous versus homogeneous ensembles, the nature of
quantum states, entropy, irreversibility, entanglement,
correlation, spontaneous decoherence, and related conceptual
issues; we are currently actively working at extracting journal
articles on this unpublished material;
\item the last chapters introduce for the first time in the physics literature a
nonlinear equation of motion for quantum dynamics which for all
mixed states, including far-nonequilibrium, implies a general law
of relaxation towards equilibrium along the ``steepest entropy
ascent'' (or ``maximal entropy generation'') trajectories in
quantal state-space. The resulting nonlinear dynamics is
compatible with the usual Hamiltonian Schr\"odinger-von Neumann
description of interactions and energy conservation, as well as
all the conditions listed in G.P. Beretta,
 ``Nonlinear extensions of Schr\"odinger-von Neumann quantum dynamics: a set
  of necessary conditions for
compatibility with thermodynamics,'' \emph{Mod.\ Phys.\ Lett.\ \bf
A20}, 977 (2005). In particular, for pure quantum mechanical
states it reduces to the standard unitary Schr\"odinger-von
Neumann evolution.
 Several publications based
on this thesis have refined the mathematics and disclosed further
intriguing features of the general dynamical law for irreversible
processes (see below);
\item being dated over twenty years ago, the publications based on
this thesis may escape a superficial literature search, especially
if based mainly on electronic search, because
 they are still unavailable for download in electronic
format. We hope that posting the thesis  will help to prevent
further `rediscoveries' such as that by Gheorghiu-Svirschevski
(see references below). With the same intent  to assure proper
acknowledgements of previous fundamental and pioneering work in
this area, the web site {\tt www.quantumthemodynamics.org} now
provides some reprints of these and other related publications.

\end{itemize}

\vspace{\baselineskip}

\noindent  To date, the thesis is  referred to in the following
references (for more references on ``quantum thermodynamics'' see
{\tt www.quantumthemodynamics.org}): \vspace{\baselineskip}

\noindent G.P. Beretta, ``A general nonlinear evolution equation
for irreversible conservative approach to stable equilibrium'', in
{\it Frontiers of Nonequilibrium Statistical Physics,} proceedings
of the NATO Advanced Study Institute, Santa Fe, June 1984, edited
by G.T.\ Moore and M.O. Scully, {\it NATO ASI Series B: Physics}
{\bf 135}, Plenum Press, New York, p.\ 193 (1986);
\vspace{\baselineskip}

\noindent G.P. Beretta, ``Intrinsic entropy and intrinsic
irreversibility for a single isolated constituent of matter:
broader kinematics and generalized nonlinear dynamics'', in {\it
Frontiers of Nonequilibrium Statistical Physics,} proceedings of
the NATO Advanced Study Institute, Santa Fe, June 1984, edited by
G.T.\ Moore and M.O. Scully, {\it NATO ASI Series B: Physics} {\bf
135}, Plenum Press, New York, p.\ 205 (1986);
\vspace{\baselineskip}

\noindent G.P. Beretta, ``On the relation between classical ad
quantum thermodynamic entropy'', \emph{J. Math. Phys \bf 25}, 1507
(1984); \vspace{\baselineskip}

\noindent G.P. Beretta, E.P. Gyftopoulos, J.L. Park, and G.N.
Hatsopoulos, ``Quantum thermodynamics. A new equation of motion
for a single constituent of matter'', \emph{Nuovo Cimento \bf
B82}, 169 (1984);

\vspace{\baselineskip} \noindent G.P. Beretta, E.P. Gyftopoulos,
and J.L. Park, ``Quantum thermodynamics. A new equation of motion
for a general quantum system'', \emph{Nuovo Cimento \bf B87}, 77
(1985);

\vspace{\baselineskip}

\noindent G.P. Beretta, ``Entropy and irreversibility for a single
isolated two level system: new individual quantum states and new
nonlinear equation of motion'', \emph{Int. J. Theor. Phys. \bf
24}, 119 (1985); \vspace{\baselineskip}

\noindent G.P.\ Beretta, ``Effect of irreversible atomic
relaxation on resonance fluorescence, absorption, and stimulated
emission'', \emph{Int. J. Theor. Phys. \bf 24}, 1233 (1985);

\vspace{\baselineskip}

\noindent G.P. Beretta, ``A theorem on Lyapunov stability for
dynamical systems and a conjecture on a property of entropy'',
\emph{J. Math. Phys \bf 27}, 305 (1986); the conjecture therein
was later found proved in F. Hiai, M. Ohya, and M. Tsukada,
\emph{Pac. J. Math. \bf 96}, 99 (1981). \vspace{\baselineskip}

\noindent G.P. Beretta, ``Steepest entropy ascent  in quantum
thermodynamics," in {\it The Physics of Phase Space,} edited by
Y.\,S.\ Kim and W.\,W.\ Zachary, {\it Lecture Notes in Physics}
{\bf 278}, Springer-Verlag, New York,  p.\ 441  (1986);
\vspace{\baselineskip}

\noindent G.P. Beretta, ``Quantum thermodynamics of
nonequilibrium. Onsager reciprocity and dispersion-dissipation
relations'', \emph{Found. Phys. \bf 17}, 365 (1987);
\vspace{\baselineskip}

\noindent S. Gheorghiu-Svirschevski, Addendum to ``Nonlinear
quantum evolution with maximal entropy production", \emph{Phys.
Rev. \bf A63}, 054102 (2001), see also \emph{Phys. Rev. \bf A63},
022105 (2001); \vspace{\baselineskip}

\noindent G.P. Beretta, ``Maximal-entropy-generation-rate
nonlinear quantum dynamics compatible with second law,
reciprocity, fluctuation--dissipation, and time--energy
uncertainty relations'', arXiv:quant-ph/0112046 (2001);
\vspace{\baselineskip}

\noindent G.P. Beretta, ``A nonlinear model dynamics for
closed-system, constrained, maximal-entropy-generation relaxation
by energy redistribution'', arXiv:quant-ph/0501178 (2005);

\vspace{\baselineskip}

\noindent  E.P. Gyftopoulos and G.P. Beretta, ``What is the second
law of thermodynamics and are there any limits to its validity?'',
arXiv:quant-ph/0507187 (2005).


\newpage
\setcounter{page}{2}
\renewcommand{\thepage}{\roman{page}}

\subsection*{} \label{abst}
\addcontentsline{toc}{chapter}{ABSTRACT}

\begin{center}
ON THE GENERAL EQUATION OF MOTION OF QUANTUM THERMODYNAMICS

\vspace{\baselineskip}

AND

\vspace{\baselineskip}

THE DISTINCTION BETWEEN QUANTAL AND NONQUANTAL UNCERTAINTIES

\vspace{\baselineskip}

by \\
Gian Paolo Beretta

\vspace{\baselineskip}

Submitted to the Massachusetts Institute of Technology \\
on August 1, 1981 in partial fulfillment of the \\
requirements for the degree of Doctor of Science

\vspace{\baselineskip}

\vspace{\baselineskip}

\vspace{\baselineskip}

\vspace{\baselineskip}

ABSTRACT

\end{center}

\vspace{\baselineskip}

A general quantum theory encompassing Mechanics, Thermodynamics
and irreversible dynamics is presented in two parts.

\noindent The first part is concerned exclusively with the
description of the states of any individual physical system. It is
based on a new nonlinear quantum equation of motion, which reduces
to the Schr\"odinger equation of motion of conventional quantum
dynamics only under special conditions. It accounts for the
implications of the laws of Thermodynamics as well as for
irreversible phenomena, such as the natural tendency of an
isolated system to transit from any non-equilibrium state to an
equilibrium state of higher entropy. Conversely, the laws of
Thermodynamics and irreversibility emerge as manifestations of the
fundamental quantum dynamical behaviour of the elementary
constituents of any material system. We call this part Quantum
Thermodynamics.

\noindent The second part of the theory, which contains the first
as a special case, is concerned with the description of stochastic
distributions of states in an ensemble of identical physical
systems each of which individually obeys the laws of Quantum
Thermodynamics. It is based on a new measure-theoretic description
of ensembles. It accounts unambiguously for the essential
distinction between two types of uncertainties that are generally
present in an ensemble, namely, quantal uncertainties due to the
inherent quantal nature of the states of each individual member
system and nonquantal uncertainties due to the stochastic
distribution of states. We call this part Quantum Statistical
Thermodynamics.

\vspace{\baselineskip}

\noindent Thesis Supervisor: \begin{tabular}[t]{l}
Elias P. Gyftopoulos \\
Ford Professor
\end{tabular}

\vspace{\baselineskip}

\noindent \copyright Massachusetts Institute of Technology, August
1981

\pagestyle{plain}
\newpage
 \addcontentsline{toc}{chapter}{TABLE OF CONTENTS}
\tableofcontents

\chapter*{Acknowledgements} \label{ackn}
\addcontentsline{toc}{chapter}{ACKNOWLEDGEMENTS}

It is a pleasure to acknowledge my indebtedness to those who have
helped me in this work with many stimulating discussions and
conversations.

My interest in the problems of the unification of Thermodynamics
and Quantum Mechanics was stimulated by innumerable conversations
with Professor Elias P. Gyftopoulos, M.I.T., throughout the
1979-80 academic year. My determination to address the
irreversibility problem was originally stimulated by a lecture
that Professor James L. Park, Washington State University, gave at
M.I.T. in April 1979.

During the 1980-81 academic year I enjoyed many conversations on
these problems with Professor Park (visiting M.I.T.) and Professor
Gyftopoulos, my thesis supervisor. In several occasions I also
enjoyed the enthusiastic and stimulating participation of Doctor
George N. Hatsopoulos, president of the Thermo Electron
Corporation.

Often, in these meetings, I have put on trial my results and
extracted, from the subsequent discussions, ideas for further
development. Specifically, I followed this patters in refining and
generalizing the equation of motion that is proposed here, the
first embryo of which I had put on trial in early February 1981.

I am especially grateful to Professor Gyftopoulos also for his
thorough critical revision of several drafts of the manuscript.

I am pleased to thank Professor James C. Keck, M.I.T., for his
encouragement, support and critical help, and Professor Gian Carlo
Rota, M.I.T., for his participation in my thesis committee.

This work was supported in part by a grant of the Thermo Electron
Corporation.

\vspace{\baselineskip} \noindent Gian Paolo Beretta

 \noindent Cambridge, August 1, 1981.
\newpage

\printindex
\newpage
\setcounter{page}{1}
\renewcommand{\thepage}{\arabic{page}}

\chapter{INTRODUCTION} \label{cap:1}

This dissertation has two objectives.

The first objective is to present  a quantum theory of nature
encompassing Mechanics, Thermodynamics and irreversible dynamics.
This theory is concerned exclusively with the description of the
states of any individual physical system. The objective is
accomplished by generalizing the currently accepted formulation of
quantum theory both as regards the description of the allowed
states of an individual physical system and as regards the
description of their causal evolution. This theory will be called
Quantum Thermodynamics.

The second objective is to present  an unambiguous statistical
theory superimposed to Quantum Thermodynamics. This statistical
theory is concerned with the description of stochastic
distributions of states in an ensemble of identical physical
systems each of which individually obeys the laws of Quantum
Thermodynamics. The objective is accomplished by adopting a new
measure-theoretic description of ensembles. This theory will be
called Quantum Statistical Thermodynamics.

In the background is the work of  G.N. Hatsopoulos and E.P.
Gyftopoulos (1976). They assumed the laws of Thermodynamics as
complementary to, consistent with, and independent of, the laws of
Mechanics and formulated a unified quantum theory which
encompasses within a single structure both Thermodynamics and
Mechanics. This was accomplished by adjoining to the fundamental
postulates of conventional quantum theory an additional
independent postulate reflecting the essence of the laws of
Thermodynamics. Three of their major conclusions can be stated as
follows:

\begin{quote}\begin{description}
    \item[(a) ] There exist quantum states  of an individual
    physical system that are more general than those considered in
    conventional Quantum Mechanics.
    \item[(b) ] The presently known law  of causal evolution of
    quantum dynamics, based on the Schr\"odinger equation of
    motion, is incomplete. Not only it cannot describe
    irreversible processes but it cannot even describe some known
    reversible processes.
    \item[(c) ] The distinction between states of an individual
physical system and distributions of states in an ensemble of
identical systems is essential. The unified quantum theory is
concerned only with the states of an individual physical system.
An unambiguous quantum statistical theory concerned with the
distributions of states in an ensemble is presently lacking. Such
a theory will be unambiguous if, and only if, it can uniquely
account for the statistical composition of each ensemble, i.e. it
unambiguously identifies both the spectrum of individual states
represented in the ensemble and their relative population.
\end{description}\end{quote}

\noindent Conclusion (a) will be retained  in this dissertation as
a fundamental premise. Conclusions (b) and (c) motivate the two
objectives of the dissertation.

As regards the first objective, Quantum  Thermodynamics resolves
the incompleteness of the present law of causal evolution. This is
accomplished by postulating a new nonlinear equation of motion.
The proposed equation of motion accounts for the implications of
the laws of Thermodynamics, such as the existence of a unique
stable equilibrium state of an individual system for any allowed
set of initial values of its constants of the motion, as well as
for irreversible phenomena, such as the tendency of an isolated
system to transit from any initial non-equilibrium state to an
equilibrium state of higher entropy. Moreover, it reduces to the
Schr\"odinger equation of motion for a particular class of states.
The resulting dynamical postulate of Quantum Thermodynamics
replaces the dynamical and the additional independent postulates
of the Hatsopoulos-Gyftopoulos unified quantum theory. Another
novel characteristic of Quantum Thermodynamics is a substantial
enlargement of the class of physical observable representatives to
include nonlinear functionals of the state operator.

As regards the second objective, Quantum  Statistical
Thermodynamics resolves the present lack of an unambiguous quantum
statistical theory. This is accomplished by assigning to each
ensemble a measure-theoretic descriptor, called the
statistical-weight measure, which represents the distribution of
individual quantum thermodynamical states in the ensemble. Every
statistical-weight measure admits of a unique spectral resolution
into irreducible measures, called Dirac measures. Since each Dirac
measure identifies a single quantum thermodynamical state, the
corresponding coefficient in the spectral resolution of the
statistical-weight measure represents the relative population, in
the ensemble, of systems being in that individual quantum
thermodynamical state. The unambiguity of Quantum Statistical
Thermodynamics is ensured by the uniqueness of the spectral
resolution of any statistical-weight measure. Consistently,
Quantum Statistical Thermodynamics reduces to Quantum
Thermodynamics for the class of statistical-weight measures that
are themselves Dirac measures and correspond to ensembles composed
of systems each of which is in exactly the same individual quantum
thermodynamical state as all the other systems. Quantum
Statistical Thermodynamics contains Quantum Thermodynamics in the
same way as Classical Statistical Mechanics contains Classical
Mechanics.

\vspace{\baselineskip}

\noindent The dissertation is structured as follows:

\vspace{\baselineskip}

\noindent Chapter \ref{cap:2} presents  the background
considerations motivating the present work, including: an
inconsistency of the currently prevailing interpretation of the
physical significance of Thermodynamics, the elimination of such
inconsistency as achieved in the unified quantum theory of
Mechanics and Thermodynamics proposed by Hatsopoulos and
Gyftopoulos, two fundamental problems arising within that theory
which are presently open and are addressed in the present work,
and the premises from which we proceed to resolve the two
fundamental problems.

\vspace{\baselineskip}

\noindent Chapter \ref{cap:3} presents  a general discussion on
the concept of state of an individual system, including: a
re-examination of its conceptual significance independently of the
current formulation of quantum theory, the identification of the
sources of ambiguity in the formulation of conventional Quantum
Statistical Mechanics, and the conditions that a statistical
theory must satisfy in order to be unambiguous. This discussion is
based on the definitions of homogeneous preparations of a physical
system, the rule of correspondence between the concept of
individual state and that of homogeneous preparation, and the
distinction between a physical and a statistical theory of nature.

\vspace{\baselineskip}

\noindent Chapters \ref{cap:4} and  \ref{cap:5} present the
general quantum theory that we propose, including: the postulates
of Quantum Thermodynamics, the postulates of Quantum Statistical
Thermodynamics, and major theorems.

\vspace{\baselineskip}

\noindent Chapter \ref{cap:6} concludes  with a summary and
recommendations for future studies.

\chapter{BACKGROUND} \label{cap:2}

This Chapter presents the background  considerations motivating
the present work. In Section \ref{sez:2.1}, the currently
prevailing understanding of the physical significance of
Thermodynamics is examined and shown to be inconsistent. It is
then observed that within a recently proposed unified quantum
theory of Mechanics and Thermodynamics, this inconsistency is
eliminated. However, other important problems remain to be solved.
In Section \ref{sez:2.2}, the two fundamental problems addressed
in this dissertation are outlined. Our approach to the solution of
these problems is based on three independent premises that are
explicitly stated at the end of the Chapter.

\section{PHYSICAL SIGNIFICANCE OF THERMODYNAMICS} \label{sez:2.1}

The prevailing interpretation of  the physical significance of
Thermodynamics, which is almost invariably proposed today, can be
traced back to the works of J.C. Maxwell (1859), L.E. Boltzmann
(1871) and J.W. Gibbs (1902), though the modern version has been
filtered through the more recent ideas of Information Theory.

Thermodynamics is considered as a  statistical theory of complex
macroscopic systems. According to P. Glansdorff and I. Prigogine
(1971), ``the great importance of thermodynamic ... methods is
that they provide us with a 'reduced description', a 'simplified
language' with which to describe macroscopic systems." Thus,
Thermodynamics would be a particularly effective set of methods to
treat complex macroscopic systems, whose mechanical description
(``the" fundamental description) is practically too complicated
and conceptually of little interest because the initial conditions
are hardly reproducible. Moreover, for systems with a sufficiently
large number of degrees of freedom, i.e. for complex macroscopic
systems, the laws of Thermodynamics would be ``derivable" from the
fundamental mechanical description.

\subsection{Prevailing View: Statistical Interpretation} \label{sotsez:2.1.1}

According to the statistical  (or information-theoretic)
interpretation, the ``thermodynamic state" of a system is
conceived of as the best description of the state of knowledge of
an observer possessing only partial information about the ``actual
state" in which the observed system actually is. Especially for
macroscopic systems, there are so many possible ``actual states",
that it would be fruitless, in addition to practically impossible,
to know which one is actually realized.

The state of knowledge of the observer  is represented by
probabilities $ p_i $ assigned to each possible ``actual state".
The assignment of such probabilities is based on an
\emph{additional postulate} adopted as a complement to the laws of
Mechanics. The currently prevailing additional postulate is the
\emph{maximum ``entropy" principle}, cf. E.T. Jaynes (1957). For
reasons discussed in Chapter 4, we will call it the maximum
``statistical uncertainty" principle. According to such postulate,
the probabilities $ p_i $ should be assigned so as to maximize the
\emph{statistical uncertainty} of the observer consistently with
whatever information is available to him. This method guarantees
the least biased assignment. The indicator of statistical
uncertainty, that is to be maximized, is that established by C.E.
Shannon (1948) in the context of Information Theory, i.e.

\be I = - \, \sum_i p_i \ln p_i \, . \label{2.1} \ee

This indicator $ I $ is commonly called  the ``entropy" and
identified with the concept of entropy as implied by the laws of
Thermodynamics. The historical background of this statistical or
information-theoretic interpretation of Thermodynamics, which
includes the works of Maxwell, Boltzmann, Gibbs and Shannon, has
been recently reviewed by Jaynes (1978).

The innumerable successes of the mathematical  formula for the
indicator of statistical uncertainty, or the ``entropy", have been
interpreted as a guarantee of correctness of the attached
statistical interpretation. However, E.P. Gyftopoulos and G.N.
Hatsopoulos (1980) have shown that such an interpretation leads to
an unacceptable inconsistency.

\subsection{Inconsistency of the Statistical Interpretation} \label{sotsez:2.1.2}

The inconsistency stems from the incompatibility  of the premise
that the thermodynamic state is a subjective characteristic of a
partially informed observer, with a large body of experimental
facts.

For example, it is an experimental fact  that any observer can
extract a net amount of work from a system in a non-equilibrium
state without leaving other net effects external to the system.
This the observer can do even if he is totally ignorant about the
``actual states" of the system. This we do in fact when we extract
work from a charged battery and yet ignore the ``actual states" of
the battery. The premise that the thermodynamic state is
subjective, and a function of the knowledge of the observer, would
instead lead to conclude, inconsistently with empirical
experience, that no net amount of work can be extracted by a
totally ignorant observer.

Again, it is an experimental fact that an  isolated system in a
non-equilibrium state generally proceeds irreversibly towards an
equilibrium state. Such a process occurs in the system, not in the
mind of the observer. Indeed, the amount of work that can be
extracted from the system at the end of such a process has
diminished, and there is nothing any observer can do to restore it
without leaving net effects external to the system.

These examples belong to a body of experimental  facts that are
consistently regularized by the laws of Thermodynamics but are
impossible to explain when its statistical interpretation is
assumed. Hence, the statistical interpretation of the physical
significance of Thermodynamics is inconsistent.

\subsection{Elimination of the Inconsistency} \label{sotsez:2.1.3}

G.N. Hatsopoulos and E.P. Gyftopoulos (1976) have  indicated that
the inconsistency can be eliminated by considering the laws of
Thermodynamics as independent of and complementary to the laws of
Mechanics. They have developed an unified quantum theory
encompassing within a single structure both Mechanics and
Thermodynamics, in which the laws of Thermodynamics emerge as
objective manifestations of the inherent quantal characteristics
of any material system. Observers and their state of knowledge
play no role in the theory. Hence, the inconsistency emerging from
the premise that the thermodynamic state is subjective is
eliminated.

The Hatsopoulos-Gyftopoulos unified  quantum theory was developed
by adjoining to the fundamental postulates of conventional quantum
theory, an additional independent postulate reflecting the essence
of the laws of Thermodynamics.

A fundamental implication of the unified  quantum theory is that
the class of quantum states of any individual physical system is
broader than the class of quantum states considered in
conventional Quantum Mechanics. Such a broader class of quantum
states includes equilibrium and non-equilibrium states. The
maximum amount of work that can be extracted from any system
without leaving other net effects external to the system is solely
a function of the quantum state of the system. If the system is in
a stable equilibrium state, then no work can be extracted without
leaving net effects external to the system. This is a statement of
the impossibility of a perpetual motion machine of the second
kind. If the system is in a non-equilibrium state, then the
maximum amount of work that can be extracted is nonzero. This
explains the body of experimental facts considered in the last
Section.

The inconsistency is eliminated,  however, two major problems
remain open and motivate the present work. First, the presently
known equation of motion of quantum dynamics is incomplete,
because it cannot explain irreversible processes in addition to
some known important reversible processes. Second, an unambiguous
quantum statistical theory for the description of distributions of
states in an ensemble of identical systems is presently lacking.

\section{OPEN PROBLEMS AND PRESENT PREMISES} \label{sez:2.2}

\subsection{Lack of a Complete Equation of Motion} \label{sotsez:2.2.1}

It is a well known fact of today's  physics that any known
\emph{fundamental} description of the dynamical behavior of any
physical system (complex or simple, macroscopic or microscopic) is
reversible. This is a characteristic feature of classical
Hamiltonian dynamics (i.e., the Hamilton equations of motion) as
well as quantum Hamiltonian dynamics (i.e., the Schr\"odinger
equation of motion). However, it is also well known that
irreversibility dominates the empirical world. Hence, the
Hamiltonian dynamical description of physical systems is
incompatible with irreversible physical reality. Several
rationalizations of this paradox have been proposed during the
last century. All are based on the assumption that irreversibility
is \emph{not} an objective phenomenon inherent in the nature of
material systems. For example, some view irreversibility as an
``antropomorphic" concept expressing the ever increasing
obsolescence of past information (cf. Jaynes (1957)). However, if
the laws of Thermodynamics are understood as objective fundamental
laws of nature, then these rationalizations cannot be conclusive.

Within the Hatsopoulos-Gyftopoulos  unified quantum theory the
paradox acquired a novel significance. A general theory
encompassing the laws of Mechanics and Thermodynamics should
account for all known reversible and irreversible processes. If
the known equations of motion cannot account for some known
process, then they are incomplete. For example, the Schr\"odinger
equation of motion is incomplete. We will return to this point in
Chapter \ref{cap:5}.

To resolve the incompleteness of the  present dynamical theory, we
will proceed from the premise that irreversibility should be
explained as a fundamental physical phenomenon due to the inherent
dynamical nature of the elementary constituents of any material
system and the premise that the laws of Thermodynamics should
follow as theorems of a complete dynamical theory.

\subsection{Lack of an Unambiguous  Quantum Statistical Theory} \label{sotsez:2.2.2}

The distinction between \emph{state} of  an individual physical
system and stochastic \emph{distribution of states} in an ensemble
of identical physical systems is essential to the
Hatsopoulos-Gyftopoulos unified quantum theory of Mechanics and
Thermodynamics. The theory is in fact concerned only with a
fundamental description of the states of any individual physical
system.

A quantum statistical theory capable of  describing unambiguously
distributions of states in an ensemble of identical systems is
presently not available. Yet, such a theory would be of
theoretical importance to clarify the physical significance of the
unified quantum theory and of practical importance to provide the
ground for a description of situations in which maximal
theoretical information on the state of a system is not available
and the methods of statistical inference must be used.

A statistical theory will be unambiguous  if, and only if, it can
uniquely account for the statistical composition of each ensemble,
i.e. it unambiguously identifies both the spectrum of states
represented in the ensemble and their relative population. The
physical importance of this requirement can be clarified as
follows. A well known striking conclusion of conventional quantum
theory, which has been enhanced by the unified quantum theory, is
the existence of uncertainties, that we call \emph{quantal
uncertainties}, which are inherent in the nature of the states of
any individual physical system. In addition to this type of
uncertainties, a quantum statistical theory must also take into
account an independent type of uncertainties, that we call
\emph{nonquantal uncertainties}, originating from the fact that in
general the member systems of an ensemble are stochastically
distributed over a range of different individual states. The
theory is unambiguous if and only if it provides a description of
ensembles which maintains the essential distinction between
quantal and nonquantal uncertainties. For example, we will see
that conventional Quantum Statistical Mechanics does not satisfy
this requirement and is therefore ambiguous.

Implicit in the foregoing discussion is  the premise that the
concept of state of an individual physical system is fundamental
in a theory of nature. In other words, it should be always
possible to think of any individual physical system as being in
some definite state. To resolve the present lack of an unambiguous
statistical quantum theory, we will proceed in Chapter \ref{cap:3}
by capitalizing on this premise.

\subsection{Premises of the Present Work} \label{sotsez:2.2.3}

Our objective in this  dissertation is to solve the two
fundamental problems just outlined, while retaining the successful
achievements of the Hatsopoulos-Gyftopoulos unified quantum theory
of Mechanics and Thermodynamics. The following statements
summarize the premises of our approach:

\begin{quote}\begin{description}
    \item[(a) ] The enlargement  of the class of quantum states of
    any individual physical system, which has been introduced in
    the unified quantum theory, should be retained, since it is
    essential to explain consistently the laws of Thermodynamics
    as fundamental laws of nature.
    \item[(b) ] The concept  of state of an individual system is
    fundamental. It should be always possible to think of any
    individual system as being in some definite state. A quantum
    statistical theory should include such a concept without
    ambiguities.
    \item[(c) ] The equation  of motion of Hamiltonian dynamics
    (i.e., the Schr\"odinger equation) is incomplete; therefore, a
    new complete equation of motion is needed. A complete
    dynamical theory should explain irreversibility as a
    fundamental physical phenomenon and should entail the laws of
    Thermodynamics as theorems.
\end{description}\end{quote}

\noindent From these premises  we proceed to develop a general
quantum theory in which the two problems just outlined have a
consistent solution.

\chapter{CONCEPT OF STATE OF AN INDIVIDUAL SYSTEM} \label{cap:3}
\vskip-4mm
 This Chapter presents a general discussion  on the
concept of state of an individual physical system. In classical
theory, this concept is well understood and established, and it
plays a fundamental role, since every system is thought of as
always being in some definite individual state. We will
hypothesize that the concept can play a fundamental role even in
quantum theory. We will see that within the formulation of
conventional Quantum Statistical Mechanics, the concept of state
of an individual system is ambiguously represented. Therefore, we
shall conclude that conventional quantum theory is incomplete,
since it lacks of a statistical quantum theory representing
unambiguously this fundamental concept.

That in conventional Quantum Statistical  Mechanics the concept of
individual state is ambiguously represented, has been clearly
shown by J.L. Park (1968a) (cf. Section \ref{sotsez:3.2.2}).
However, Park concluded: ``Thus the concept of individual quantum
state is fraught with ambiguity and should therefore be avoided in
serious philosophical inquiries concerning the nature of quantum
theory ... A quantum system should be regarded as never being in
any physical state." On the contrary, here we take the position
that the concept of individual physical state is fundamental. From
this premise, we conclude that what is fraught with ambiguity is
not the concept of individual state, but rather the current
mathematical formulation of Quantum Statistical Mechanics.
Therefore, a serious reformulation of quantum theory is in order.

To pursue this objective, we proceed as  follows. We re-examine
the concept of state of an individual system independently of the
formulations of either conventional quantum theory or the
Hatsopoulos-Gyftopoulos unified quantum theory. In the light of
this re-examination, we study the physical and statistical
formulations of both classical theory and conventional quantum
theory and review the ambiguity of conventional Quantum
Statistical Mechanics. Finally, we identify necessary conditions
that a statistical theory must satisfy in order to achieve an
unambiguous representation of the concept of state of an
individual system. Using these conditions, we shall be able to
show (Chapter \ref{cap:4}) that the statistical part of the
general quantum theory that we propose is unambiguous.

\section{RULES OF CORRESPONDENCE AND\\ DEFINITION OF PARADIGM} \label{sez:3.1}

A scientific theory of nature comprehends three realms:

\vspace{\baselineskip}

\noindent - the realm of empirical experience (perceptions)

\noindent - the realm of theoretical  constructs (conceptions)

\noindent - the realm of mathematical  descriptions
(representations).

\vspace{\baselineskip}

\noindent The essence of the theory  resides in the \emph{links}
that it establishes between elements of these three realms.

The links between perceptions and  conceptions have been called
\emph{rules of correspondence} by H. Margenau (1950): ``The rules
of correspondence ... are not eternally grounded in the nature of
things, nor are they immediately suggested by sensory experience;
they are important parts of every theory of nature and receive
their validity from the consistency, the internal neatness and
success of the entire explanatory scheme."

The rules of correspondence are normally defined through the
mathematical formulation of the theory. Practitioners do not work
with perceptions, conceptions and rules of correspondence
expressed by verbose locutions. Rather, they work with data,
mathematical entities and mathematical relations, i.e. with
\emph{mathematical descriptions} of perceptions, conceptions and
rules of  correspondence. The mathematical formulation is
essential in giving structure to the theory; however, it is not
generally unique. A theory might admit of several different, but
mathematically equivalent, formulations.

A set of postulates giving (in their unity)  structure to a
theory, by linking theoretical constructs and rules of
correspondence to mathematical descriptors and mathematical
relations, can be called the \emph{paradigm} of the theory. This
terminology has been used by T.S. Kuhn (1962), in his book on The
Structure of Scientific Revolutions.

The following two Sections are intended  to present our general,
non-mathematical re-examination of the concept of state of an
individual system, based on the definition of the fundamental
rules   of correspondence between \emph{states, homogeneous
ensembles and homogeneous preparations}.

\subsection{States and Homogeneous Ensembles} \label{sotsez:3.1.1}

As discussed by J.L. Park (1968a),  the dominant theme of quantum
theory is the necessity to accept that the notion of state
involves probabilistic concepts in an essential way. In quantum
theory the links between \emph{probabilities} and the empirical
world are implicitly defined by established practices of
experimental science: the construct probability is linked to the
relative frequency in an \emph{ensemble}. Thus the primary
connection of a quantum theory of nature to the empirical world is
through ensembles.

The purpose of a quantum theory is to  regularize purely
probabilistic information about the measurement results from a
real ensemble of identical systems. A \emph{real ensemble} is
either an aggregate of identically prepared systems upon which
measurements are performed or a sequence of measurements performed
on a single system identically reprepared prior to each
measurement, or a combination of both.

An important scheme for the classification  of ensembles,
especially emphasized by J. von Neumann (1932), hinges upon the
concept of ensemble homogeneity. Given an ensemble it is always
possible to conceive of it as subdivided into many subensembles.

\vspace{\baselineskip}

\noindent Definition: Homogeneous Ensembles

\begin{quote}
\begin{em}
An ensemble is said to be homogeneous  if, and only if, every
conceivable subdivision results into subensembles all identical to
the original.
\end{em}
\end{quote}

\noindent Here, two ensembles of  identical systems are said to be
\emph{identical} if, and only if, the arithmetic mean value of
data yielded by measurements of a physical observable on the
member systems of one of the two ensembles is equal to the mean
value of data yielded by measurements of the same physical
observable on the member systems of the other ensemble, for
\emph{all conceivable} physical observables.

Since, by definition, there exist no  subdivisions of a
homogeneous ensemble into \emph{different} subensembles, it
follows that each individual member system of a homogeneous
ensemble has exactly the same \emph{intrinsic characteristics} as
any other member. For, if all member systems did not have the same
intrinsic characteristics, then it would be possible to conceive
of subdivisions of the ensemble into subensembles that are not
identical.

The whole of these intrinsic characteristics,  shared by all the
individual members systems of a homogeneous ensemble, defines the
concept of \emph{state} of an individual system. In short, a
homogeneous ensemble is a real aggregate of identical systems each
individually in the same state. Thus, the concept of state
acquires physical meaning in its reference to the \emph{individual
system}, though its empirical correspondent is a homogeneous
ensemble of identical systems. In fact, we can now express the
rule of correspondence linking the physical concept of state to
the empirical world:

\begin{quote}
The empirical correspondent of the  theoretical concept of state
of an individual system is a particular kind of ensemble: the
\emph{homogeneous ensemble}.
\end{quote}

\noindent The power of this rule of  correspondence is that it
holds with the same fundamental meaning whether the concept of
state does or does not involve probabilistic concepts.

\subsection{Preparations and Ensembles} \label{sotsez:3.1.2}

The act of generating an ensemble  consists of the repeated
application, on the member systems, of a list of operations called
an ensemble preparation scheme or, simply, a \emph{preparation}.
By definition, a preparation must be \emph{reproducible} in that
it must be identically applied to all the members that are to form
the ensemble.

It is noteworthy that the act of  generating an ensemble, i.e. the
concept of \emph{preparation}, does not necessarily involve an
interaction with the member systems of the ensemble itself. It is
only the following \emph{measurement} act which must necessarily
involve an interaction. The concepts of preparation and
measurement must not be confused (cf. H. Margenau and J.L. Park
(1973)). A measurement act is an ``operation performed on a system
for the purpose of obtaining a numerical value which can, by
virtue of the chosen experimental arrangement, be assigned to some
definite, nameable observable." A preparation scheme is instead
any set of operations selected for the purpose of obtaining a real
ensemble of identical systems. Thus a preparation is a scheme to
``accept" systems upon which to perform measurements.

Deliberately, the concept of  preparation as just defined is quite
broad and unrestricted. In general, not all preparations generate
homogeneous ensembles. If a preparation generates a homogeneous
ensemble, it will be called a \emph{homogeneous preparation},
otherwise, it will be called a \emph{heterogeneous preparation}.

It is useful to define a shorthand  notation to represent the
notion of \emph{statistical composition} of two, or more,
preparation schemes. Let $ \Pi_1 $ and $ \Pi_2 $ represent two
preparation schemes for a physical system. If the system is
prepared according to scheme $ \Pi_1 $ with statistical weight $
w_1 $ or according to scheme $ \Pi_2 $ with statistical weight $
w_2 $ (where $ w_1 + w_2 = 1 $ and $ w_l, w_2 > 0 $), then we
indicate the resulting composite preparation scheme $ \Pi $ by
writing

\bd \Pi = w_1 \Pi_1 + w_2 \Pi_2 \, , \ed

\noindent where this expression  must be understood exclusively as
a shorthand symbolic notation having no other meaning than that
just cited. With this notation, and by the definition of
homogeneous ensemble, we can give an alternative definition of
homogeneous preparation.

\vspace{\baselineskip}

 \noindent Definition: Homogeneous
Preparations

\begin{quote}
A preparation scheme is said to  be homogeneous if, and only if,
it cannot result from the statistical composition of different
preparation schemes, i.e. if, and only if, any conceivable
decomposition of the type

\bd \Pi = w_1 \Pi_1 + w_2 \Pi_2 \quad \mbox{ with } w_1 , \, w_2 >
0 \ed

implies that $ \Pi_1 = \Pi_2 = \Pi $.
\end{quote}

As seen in the last Section, to  the theoretical concept of
\emph{state} of an individual system is linked the empirical
concept of \emph{homogeneous ensemble}, which, in turn, has been
linked here to the theoretical concept of \emph{homogeneous
preparation}. Thus the two theoretical concepts of \emph{state}
and \emph{homogeneous preparation} are linked through their
correspondence with the \emph{homogeneous ensemble}. If the
concept of state of an individual system is fundamental to the
particular theory of nature under consideration, then so is the
concept of homogeneous preparation. The two concepts are linked
one-to-one regardless of the particular mathematical formulation
of the theory. This is the conclusion of our re-examination of the
concept of state of an individual system. The foregoing analysis
has followed closely the analysis of J.L. Park (1968a). However,
our conclusion is different, since, as we already discussed, our
premise is to consider the concept of individual state as
fundamental, regardless of the current formulation of quantum
theory.

The importance of the notion of  homogeneity, especially in the
framework of quantum theory, has been emphasized by several
authors, starting    with    J. von Neumann  (1932). G. N.
Hatsopoulos   and E. P. Gyftopoulos (1976) have proposed a unified
quantum theory of Mechanics and Thermodynamics requiring such a
notion in a fundamental way. Their theory is concerned exclusively
with a description of the homogeneous preparations, i.e. the
individual states, of a physical system. The class of individual
quantum states which are considered in their unified quantum
theory is broader than the class of states considered by von
Neumann. To underline this distinction, they adopted the term
\emph{unambiguous} preparation instead of homogeneous preparation.
In the general quantum theory that we propose in Chapters
\ref{cap:4} and \ref{cap:5}, the Hatsopoulos-Gyftopoulos broader
class of individual quantum states is maintained;   however, also
the term homogeneous preparation is maintained, since we have
shown that the concept of homogeneity stands independently of the
particular theory of nature under consideration.

\section{PHYSICS VERSUS STATISTICAL PHYSICS} \label{sez:3.2}

\emph{Physics} is that branch of the theory of nature which is
intended to describe within a causal framework the \emph{states}
of any physical system. Equivalently, the objective of  physics is
a description of the \emph{homogeneous preparations} of a general
physical system. To the concept of state, the physical theory
generally associates a \emph{mathematical descriptor}, usually an
element of a set. The same mathematical entity is thus also
associated with the corresponding homogeneous preparation and the
corresponding homogeneous ensemble.

\emph{Statistical physics} is  that branch of the theory of nature
which is intended to describe within a statistical framework the
distributions of states in an ensemble of identical systems each
of which individually obeys the laws of the physical branch of the
theory. Equivalently, the objective of statistical physics is a
statistical description of \emph{all possible preparations} of a
physical system. To the concept of preparation, regardless of its
homogeneity, the statistical theory generally associates a
mathematical descriptor. The same descriptor is thus associated
with the corresponding ensemble.

The typical construction pattern  is to first formulate physics
and then superimpose a statistical theory to it. Thus, first a
physical theory of \emph{homogeneous preparations} (states) is
developed and then a statistical theory is superimposed to it for
the description of any \emph{preparation}, regardless of its
homogeneity.

This pattern is exemplified in  the next Sections, by reviewing
classical theory -- composed of the paradigms of Classical
Mechanics and Classical Statistical Mechanics -- and conventional
quantum theory -- composed of the paradigms of Quantum Mechanics
and Quantum Statistical Mechanics.

Throughout this dissertation,  the postulates    forming the
paradigms of different theories are identified by monograms of the
type

\begin{center}
PnNT: title
\end{center}

\noindent where Pn stands for  ``Postulate number n", NT stands
for the ``Name of the Theory" and the title indicates the concept
which is being given a mathematical description.

\subsection{Classical Mechanics vs Classical Statistical Mechanics} \label{sotsez:3.2.1}

The following review serves to  clarify the role that the notions
introduced in the foregoing discussion play in the classical
theory of nature. It is presented here because this is the
\emph{only} complete (i.e., physical and statistical) theory of
nature which is free of ambiguities. Its failure is not due to
logical inconsistencies within the theory itself, but rather to
its inability to regularize a class of empirical phenomena.

\vspace{\baselineskip}

\vspace{\baselineskip}

\noindent CLASSICAL MECHANICS

\vspace{\baselineskip}

\noindent P1CM: Systems

\begin{quote}
To every physical \emph{system}  there corresponds a real
phase-space $ \Omega $ whose elements are points with coordinates
indicated by $ (\sot{q}, \sot{p}) $.
\end{quote}

\noindent P2CM: Homogeneous Preparations

\begin{quote}
To every \emph{homogeneous preparation}  scheme $ \Pi $ for a
system, there corresponds a point $ (\sot{q}, \sot{p}) $ in the
phase-space $ \Omega $ of the system.
\end{quote}

\break

\noindent P3CM: Physical Observables

\begin{quote}
Some real functions g, h, ...  defined on $ \Omega $ correspond to
\emph{physical observables} of the system. Given an ensemble of
identical systems prepared according to the homogeneous scheme $
\Pi $, with corresponding point $ (\sot{q}, \sot{p}) $ in
phase-space, the arithmetic \emph{mean value}  $ \overline{g} $ of
data yielded by measurements of the observable $ g $ is given by
the value of the function $ g $ at the point $ (\sot{q}, \sot{p})
$, namely, \bd \overline{g} = g (\sot{q}, \sot{p}) \, . \ed
\end{quote}

\noindent P4CM: States

\begin{quote}
Every individual system is always  in a \emph{state} described by
some point $ (\sot{q}, \sot{p}) $ in the corresponding
phase-space.
\end{quote}

\noindent P5CM: Causal Evolution

\begin{quote}
For every physical system there exists  a function $ h $ (the
Hamiltonian function), defined on the phase-space of the system,
which determines the \emph{causal evolution} of the state
descriptor $ (\sot{q}, \sot{p}) $ via the following law
(Hamilton's equations of motion) \be \fr{d \sot{q}}{d t} =
\fr{\partial h}{\partial \sot{p}} \quad \fr{d \sot{p}}{d t} = - \,
\fr{\partial h}{\partial \sot{q}} \, . \label{3.1} \ee
\end{quote}

\noindent This formulation of Classical  Mechanics is
non-orthodox, especially as regards postulates P2CM and P3CM. It
has been adopted here in order to maintain a parallel structure
with the formulations of other theories considered in the
dissertation, and to exemplify the role played by the concept of
homogeneous preparation in this well established theory of nature.

\vspace{\baselineskip}

\vspace{\baselineskip}

\noindent CLASSICAL STATISTICAL MECHANICS

\vspace{\baselineskip}

\noindent P1CSM: Systems

\begin{quote}
Same as P1CM.
\end{quote}

\noindent P2CSM: Preparations

\begin{quote}
To every \emph{preparation} scheme $ \Pi $ for a system, there
corresponds a continuous, real, positive function $ f $ defined on
the phase-space $ \Omega $ of the system, satisfying the
normalization condition \be \int_{\Omega} f (\sot{q}, \sot{p}) d
\sot{q} d \sot{p} = 1 \, . \label{3.2} \ee The function $ f $ is
called the  \emph{density-of-phase function}.
\end{quote}

\break

\noindent P3CSM: Physical Observables

\begin{quote}
Some real functions $ g, h, \, ... $  defined on $ \Omega $
correspond to \emph{physical observables} of the system. Given an
ensemble of identical systems prepared according to the scheme $
\Pi $, with corresponding density-of-phase function $ f $, the
\emph{expected mean value} $ \langle  \overline{g} \rangle  $ of
measurements of the observable $ g $ is given by the following
integral functional \be \langle  \overline{g} \rangle  =
\int_{\Omega} f (\sot{q}, \sot{p}) g (\sot{q}, \sot{p}) d \sot{q}
d \sot{p} \, . \label{3.3} \ee
\end{quote}

\noindent P4CSM: States

\begin{quote}
Same as P4CM.
\end{quote}

\noindent P5CM: Causal Evolution

\begin{quote}
Same as P5CM. The corresponding  evolution of the preparation
descriptor (i.e, of the density-of-phase function $ f $) is given
by the equation (Liouville's equation) \be \fr{d f}{d t} = \{ h ,
 f \} \label{3.4} \ee where $ \{\  , \ \} $ represents the
Poisson bracket.
\end{quote}

\noindent Some theorems of Classical  Statistical Mechanics are as
follows. Our conclusion from these theorems will be that Classical
Statistical Mechanics provides an \emph{unambiguous} description
of preparations, regardless of their homogeneity.

\vspace{\baselineskip}

\noindent Th1CSM

\begin{quote}
Among   all the normalized   functions   definable   on  the
phase-space $ \Omega $, only the Dirac delta ``functions"    $
\delta_0 (\sot{q}, \sot{p}) = \delta (\sot{q} - {\sot{q}}_0)
\delta (\sot{p} - {\sot{p}}_0) $ are \emph{irreducible}, in the
sense that the equality \be \delta_0 (\sot{q}, \sot{p}) = w_1 f_1
(\sot{q}, \sot{p}) + w_2 f_2 (\sot{q}, \sot{p}) \mbox{ with }
\quad w_1 , \, w_2 > 0 \label{3.5} \ee holds if, and only if, $
f_1 = f_2 = \delta_0 $.
\end{quote}

\noindent This follows from  the fact that the Dirac delta
function has a \emph{support} (i.e., that part of the phase-space
for which the function is nonzero) which is indivisible, being a
single point $ ({\sot{q}}_0, {\sot{p}}_0) $.

\vspace{\baselineskip}

\noindent Th2CSM

\begin{quote}
To every distinct point  $ ({\sot{q}}_0, {\sot{p}}_0) $ in $
\Omega $ there corresponds a distinct Dirac delta function $
\delta (\sot{q} - {\sot{q}}_0) \delta (\sot{p} - {\sot{p}}_0) $
defined on $ \Omega $.
\end{quote}

\noindent This theorem establishes a  one-to-one correspondence
between Dirac delta functions on phase-space and the state
descriptors of Classical Mechanics, i.e. the points in
phase-space.

\vspace{\baselineskip}

\break

\noindent Definition: Homogeneous Preparations within CSM

\begin{quote}
A preparation is said to be homogeneous  if, and only if, it is
represented by a Dirac delta function.
\end{quote}

\noindent Th3CSM

\begin{quote}
Among all the preparation schemes for  a physical system, only the
\emph{homogeneous} preparations cannot be conceived of as the
result of a statistical composition of \emph{different}
preparation schemes. Moreover, each homogeneous preparation
corresponds to one and only one \emph{state} of an individual
system, and vice versa.
\end{quote}

\noindent This theorem discloses that  the \emph{rule of
correspondence} linking the concept of state of an individual
system to the concept of homogeneous preparation is indeed
reproduced within the paradigm of Classical Statistical Mechanics.

\vspace{\baselineskip}

\noindent Th4CSM

\begin{quote}
Every normalized function   $ f $  defined on $ \Omega $ can be
uniquely expressed as a ``weighted sum" of Dirac delta functions,
i.e. \be f (\sot{q} ,  \sot{p}) = \int_{\Omega} \delta (\sot{q} -
{\sot{q}}_0) \delta (\sot{p} - {\sot{p}}_0) f ({\sot{q}}_0 ,
{\sot{p}}_0) d {\sot{q}}_0 d {\sot{p}}_0 \label{3.6} \ee where the
function $ f ({\sot{q}}_0 ,  {\sot{p}}_0) $ itself represents the
``weight" of the corresponding Dirac delta function. Consequently,
every heterogeneous preparation can be \emph{uniquely resolved}
into its homogeneous component preparations.
\end{quote}

\noindent This last theorem states  explicitly that the
mathematical descriptor of a heterogeneous preparation is
\emph{uniquely} indicative of the statistical structure of the
preparation itself, i.e. of its homogeneous component preparations
and the associated statistical weights. It is therefore correct,
within the framework of this theory, to conceive of an individual
system, which has been prepared according to a heterogeneous
preparation, as being ``in" a \emph{state} corresponding to one of
the homogeneous component preparations, with a probability given
by the associated statistical weight.

To summarize, we have exemplified  the key role played by the
distinction between homogeneous and heterogeneous preparations
within the framework of the classical theory of nature. The
\emph{link} between homogeneous preparations and states of an
individual system is indeed reproduced in the theory. Moreover,
the \emph{concept} of state of an individual system is
unambiguously represented even when the preparation scheme is
heterogeneous.

\subsection{Quantum Mechanics vs Quantum Statistical Mechanics} \label{sotsez:3.2.2}

We now turn to the currently accepted  version of conventional
quantum theory -- composed of the paradigms of Quantum Mechanics
and Quantum Statistical Mechanics -- and show that, within this
formulation, the notion of state of an individual system is
ambiguously represented. The structure of the paradigms is
mathematically equivalent to the general structure outlined by
J.L. Park (1968b).

\vspace{\baselineskip}

\vspace{\baselineskip}

\break

\noindent QUANTUM MECHANICS

\vspace{\baselineskip}

\noindent P1QM: Systems

\begin{quote}
To every physical \emph{system}  there corresponds a separable,
complex Hilbert space $ \mathcal{H} $ whose elements are vectors $
\psi $. The Hilbert space of a system composed of two
distinguishable subsystems 1 and 2, with    corresponding Hilbert
spaces $ \mathcal{H} (1) $ and $ \mathcal{H} (2) $, is the direct
product Hilbert space $ \mathcal{H} (1) \otimes \mathcal{H} (2) $.
\end{quote}

\noindent P2QM: Homogeneous Preparations

\begin{quote}
To every \emph{homogeneous preparation}  scheme $ \Pi $ for a
system, there corresponds a unit-norm vector $ \psi $ in the
Hilbert space of the system. The vector $ \psi $ is called the
\emph{state vector}.
\end{quote}

\noindent P3QM: Physical Observables

\begin{quote}
Some linear, Hermitian operators  $ G, H, \, ... $   defined on $
\mathcal{H} $ correspond to \emph{physical observables} of the
system. Given an ensemble of identical systems prepared according
to the homogeneous scheme $ \Pi $, with corresponding state vector
$ \psi $, the arithmetic \emph{mean value} $ \overline{g} $ of
data yielded by measurements of the observable $ G $ is given by
the value of the following scalar product \be \overline{g} =
\langle \psi ,  G \psi \rangle  \, . \label{3.7} \ee
\end{quote}

\noindent P4QM: States

\begin{quote}
Every individual system is always  in a \emph{state} described by
some vector $ \psi $ in the corresponding Hilbert space.
\end{quote}

\noindent P5QM: Causal Evolution

\begin{quote}
For every physical system  there exists an operator $ H $ (the
Hamiltonian operator), defined on the Hilbert space of the system,
which determines the \emph{causal evolution} of the state
descriptor $ \psi $ via the following law (Schr\"odinger's
equation of motion) \be \fr{d \psi}{d t} = - \, \fr{i}{\hbar} H
\psi \, . \label{3.8} \ee
\end{quote}

\noindent The above paradigm  has been constructed so as to entail
the famous uncertainty theorem (W. Heisenberg (1927)).
Mathematically, the theorem follows immediately from the
Cauchy-Schwarz inequality applied to the scalar product $ \langle
\ , \ \rangle  $. Let the variance of measurement results for the
physical observable $ G $ be defined by

\be var (G) = \overline{({g - \overline{g}}^2)} = \langle  \psi ,
G^2 \psi \rangle  - {( \langle  \psi ,  G \psi \rangle  )}^2 \, .
\label{3.9} \ee

\noindent The uncertainty  theorem consists of the following well
known general inequality valid for any two Hermitian operators $ G
$ and $ F $ and therefore for two physical observables of the
system

\be var (G) var (F) \geq { | \langle  \psi ,  [ F ,  G ] \psi
\rangle  | }^2 \label{3.10} \ee

\noindent where $ [\ , \ ] $  is the standard commutator symbol $
([F, G] = FG - GF) $. Conversely, the theory reflects the
\emph{fundamental Heisenberg hypothesis}, namely, that
\emph{inherent in the nature of the state} of any individual
physical system there exist \emph{uncertainties}. These
uncertainties induce irreducible dispersions in the results of
measurements performed on any \emph{homogeneous} ensemble. No
dispersionless ensemble can even be \emph{conceived} within the
framework of Quantum Mechanics. This striking conclusion
constitutes one of the essential aspects of the departure of
Quantum Mechanics from Classical Mechanics. The existence of
uncertainties intimately connected with the nature of the state of
any individual material system is a universally accepted aspect of
today's physics.

\vspace{\baselineskip}

\vspace{\baselineskip}

\noindent QUANTUM STATISTICAL MECHANICS

\vspace{\baselineskip}

\noindent P1QSM: Systems

\begin{quote}
Same as P1QM.
\end{quote}

\noindent P2QSM: Preparations

\begin{quote}
To every \emph{preparation}  scheme $ \Pi $ for a system, there
corresponds a linear, Hermitian, nonnegative-definite, unit-trace
operator $ W $ defined on the Hilbert space of the system. The
operator $ W $ is called the \emph{statistical operator}.
\end{quote}

\noindent P3QSM: Physical Observables

\begin{quote}
Some linear, Hermitian operators  $ G, H, \, ... $   defined on $
\mathcal{H} $ correspond to \emph{physical observables} of the
system. Given an ensemble of identical systems prepared according
to the scheme $ \Pi $, with corresponding statistical operator $ W
$, the \emph{expected mean value} $ \langle  \overline{g} \rangle
$ of measurements of the observable $ G $ is given by the value of
the following trace functional \be \langle  \overline{g} \rangle =
\Tr (WG) \, . \label{3.11} \ee
\end{quote}

\noindent P4QSM: States

\begin{quote}
There can be no ``P4" postulate.  Saying that every individual
system is always in a \emph{state}, described by some vector $
\psi $ in $ \mathcal{H} $, leads to inconsistencies and paradoxes
(see the discussion at the end of this Section).
\end{quote}

\noindent P5QSM: Causal Evolution

\begin{quote}
Same as P5QM. The corresponding  evolution of the preparation
descriptor (i.e., of the statistical operator $ W $) is given by
the equation (von Neumann`s equation) \be \fr{d W}{d t} = - \,
\fr{i}{\hbar} [H ,  W] \, . \label{3.12} \ee
\end{quote}

\noindent It is noteworthy  that the \emph{statistical operator} $
W $ (which is also usually called the \emph{density operator} and
indicated by the symbol $ \rho $) has an essentially different
physical meaning than the \emph{state operator} defined within the
paradigm of Quantum Thermodynamics. This distinction will be
clarified in Section \ref{sotsez:4.1.1}.

Some of the theorems of  Quantum Statistical Mechanics are as
follows.

\vspace{\baselineskip}

\break

\noindent Th1QSM (J. von Neumann (1932))

\begin{quote}
Among the linear, Hermitian,  nonnegative-definite, unit-trace
operators $ W $ definable on the Hilbert space $ \mathcal{H} $,
only the projection operators $ P_{\psi} $ onto the
one-dimensional subspaces of $ \mathcal{H} $ are
\emph{irreducible}, in the sense that the equality \be P_{\psi} =
a_1 W_1 + a_2 W_2 \quad \mbox{with} \quad a_1 , \, a_2 > 0
\label{3.13} \ee holds if, and only if, $ W_1 = W_2 = P_{\psi} $.
\end{quote}

\noindent This follows from  the fact that the projection
operators $ P_{\psi} $ onto the one-dimensional subspaces of $
\mathcal{H} $ are the extreme elements of the convex set of
statistical operators on $ \mathcal{H} $.

\vspace{\baselineskip}

\noindent Th2QSM

\begin{quote}
To every distinct unit-norm  vector $ \psi $ in $ \mathcal{H} $
there corresponds a distinct projection operator $ P_{\psi} $ onto
the one-dimensional subspace of $ \mathcal{H} $ spanned by $ \psi
$.
\end{quote}

\noindent This theorem  establishes a one-to-one correspondence
between projection operators onto one-dimensional subspaces of the
Hilbert space and the state descriptors of Quantum Mechanics, i.e.
the state vectors in Hilbert space.

\vspace{\baselineskip}

\noindent Definition: Homogeneous Preparations within QSM

\begin{quote}
A preparation is said to be  homogeneous if, and only if, it is
described by a projection operator $ P_{\psi} $ onto a
one-dimensional subspace of $ \mathcal{H} $.
\end{quote}

\noindent Th3QSM

\begin{quote}
Among all the preparation schemes  for a physical system, only the
\emph{homogeneous} preparations cannot be conceived as the result
of a statistical composition of \emph{different} preparation
schemes. Moreover, each homogeneous preparation corresponds to one
and only one \emph{state} of an individual system, and vice versa.
\end{quote}

\noindent This theorem discloses  that the \emph{rule of
correspondence} linking the concept of state of an individual
system to the concept of homogeneous preparation is indeed
reproduced within the paradigm of Quantum Statistical Mechanics.

\vspace{\baselineskip}

\noindent Th4QSM (E. Schr\"odinger (1936))

\begin{quote}
Every non-idempotent statistical  operator $ W $ on $ \mathcal{H}
$ (i.e., the statistical operators that are not projection
operators onto one-dimensional subspaces of $ \mathcal{H} $) can
be expressed as a weighted sum of projection operators (onto
one-dimensional subspaces of $ \mathcal{H} $) in at least two
different ways (generally, an infinity of ways), such as \be W =
\sum_n w_n {P_{\psi}}_n = \sum_q w_q' {P_{\phi}}_q = \, ...
\label{3.14} \ee
\end{quote}

\noindent Consequence of this  theorem is that the \emph{concept}
of state of an individual system is \emph{not} unambiguously
represented within Quantum Statistical Mechanics and, therefore,
there can be no state postulate (P4QSM). In fact, if we insisted
that such a postulate be included in the paradigm of the theory, a
logical consequence of the last theorem would be that an
individual system prepared according to the heterogeneous
preparation scheme $ \Pi $, with corresponding statistical
operator $ W $, is, for example, in state $ \psi_n $ with
probability $ w_n $ and, at the same time, in state $ \phi_q $
with probability $ w_q' $. Such a system is what J.L. Park (1968a)
called a \emph{quantum monster}: a single system which is
concurrently ``in" \emph{two different} states.

This is the observation that  led Park to the conclusion quoted at
the beginning of this Chapter. We have already objected that the
above ambiguity about the concept of state of an individual system
should be charged to the underlying mathematical description
rather than to the concept itself.

To summarize, this Section  has exemplified the key role played by
the distinction between homogeneous and heterogeneous preparations
within the framework of the conventional quantum theory of nature.
The \emph{link} between homogeneous preparations and states of an
individual system is reproduced. However, the \emph{concept} of
state of an individual system is ambiguously represented when the
preparation scheme is heterogeneous.

To this point, we have  established the scientific need to
formulate a quantum statistical theory that includes the concept
of state of an individual system without ambiguities.

\subsection{Unambiguous Description of Heterogeneous Preparations} \label{sotsez:3.2.3}

By examining the paradigms  of Classical Statistical Mechanics and
Quantum Statistical Mechanics, we have learned that a mathematical
description of heterogeneous preparations must satisfy certain
conditions in order to be considered unambiguous. We summarize the
results by stating three necessary conditions which define
explicitly the notion of unambiguous mathematical description of
heterogeneous preparations.

A mathematical description of preparations is defined   as
follows:

\begin{quote}
\begin{description}
    \item[(o) ] To any  preparation scheme there corresponds an
    element of the set of mathematical descriptors.
\end{description}
\end{quote}

\noindent The three  necessary conditions are as follows. Each
condition is exemplified by indicating its realization within the
paradigm of CSM, where the set of mathematical descriptors is the
set of density-of-phase functions defined on the phase-space of
the system.

\begin{quote}
\begin{description}
    \item[(i) ] To the  concept of statistical composition of different
    preparation schemes there must correspond a mathematical \emph{rule}
    to combine their mathematical descriptors so as to obtain the mathematical
    descriptor of the composite preparation. This rule of combination must
    be such that, for every physical observable, the expected mean value
    corresponding to the composite preparation equals the statistically
    weighted sum of the expected mean values corresponding to the component
    preparations.
\end{description}\end{quote}

\noindent Within CSM, to  the statistical composition (with
weights $ w_1 $ and $ w_2 $) of  two preparations with
corresponding density-of-phase functions $ f_1 $ and $ f_2 $,
there corresponds the following rule of combination yielding the
density-of-phase function $ f_C $ of the composite preparation:

\bd f_C = w_1 f_1 + w_2 f_2 \, . \ed

It is immediate to  verify that the requirement about the expected
mean values is satisfied, i.e. that

\bd { \langle  \overline{g} \rangle  }_C = w_1 { \langle
\overline{g} \rangle  }_1 + w_2 { \langle  \overline{g} \rangle
}_2 \, . \ed

\begin{quote}\begin{description}
    \item[(ii) ] To  every homogeneous preparation there must correspond
    an \emph{irreducible} mathematical descriptor, i.e. an element of
    the set of mathematical descriptors which cannot be obtained from
    the combination of \emph{different} elements of the set. This
    implies not only that such irreducible elements must exist, but
    also that they must be at least as ``numerous" as the different
    homogeneous preparations.
\end{description}\end{quote}

\noindent Within CSM,  to every homogeneous preparation there
corresponds a Dirac delta ``function" which is indeed irreducible,
since its support is a single point in phase-space. Moreover, a
Dirac delta function can be defined in correspondence to every
point in phase space (cf. theorem Th2CSM).

\begin{quote}\begin{description}
    \item[(iii) ] To  the uniqueness of resolution of any preparation
    into its homogeneous component preparations there must correspond
    the uniqueness of resolution of each mathematical descriptor into
    a combination of its irreducible components.
\end{description}\end{quote}

\noindent Within CSM,  every density-of-phase function can be
uniquely resolved into a ``weighted sum" of irreducible Dirac
delta functions (cf. theorem Th4CSM) .

The most demanding  conditions are (ii) and (iii). Condition (ii)
reflects the need for a correct mathematical correspondent to the
concept of \emph{homogeneous preparation}. The irreducibility of
this class of preparations is essential and must therefore be
reflected in the mathematical description. In addition, the
description must be mathematically rich enough so that with every
distinct homogeneous preparation is associated at least one
distinct mathematical descriptor, thus reproducing the \emph{rule
of correspondence} between states and homogeneous preparations.
Condition (iii) reflects the need for an unambiguous
representation of the \emph{concept} of state of an individual
system prepared according to a heterogeneous preparation. The
uniqueness of resolution of a heterogeneous preparation into its
homogeneous component preparations is essential to avoid
paradoxical conclusions of the type discussed in last Section, and
must therefore be reflected in the mathematical description.

In our search for an  unambiguous formulation of a statistical
theory to be superimposed to Quantum Thermodynamics (cf. Section
\ref{sez:4.1}), three different mathematical descriptions of
heterogeneous preparations were considered. When analyzed in the
light of the conditions just defined, two were found
unsatisfactory (these attempts are instructive and are outlined in
appendix \ref{app:A}), but the third satisfied all three
conditions. This new description of preparations is proposed in
Section \ref{sotsez:4.2.1} and then adopted in Section
\ref{sotsez:4.2.2} to formulate the paradigm of Quantum
Statistical Thermodynamics.

\chapter{NEW GENERAL QUANTUM THEORY: STATES AND STATISTICS} \label{cap:4}

This Chapter presents  a general quantum theory in which the
concept of state of an individual system is unambiguously
represented and which encompasses conventional Quantum Mechanics,
Thermodynamics of stable equilibrium states and irreversible
dynamics. Section \ref{sez:4.1} presents the paradigm of the
physical part of the theory, that we call Quantum Thermodynamics.
Section \ref{sez:4.2} presents the paradigm of the statistical
part, that we call Quantum Statistical Thermodynamics.

\section{QUANTUM THERMODYNAMICS} \label{sez:4.1}

\subsection{The paradigm of Quantum Thermodynamics} \label{sotsez:4.1.1}

The paradigm of Quantum Thermodynamics is  based on the following
five postulates.

\vspace{\baselineskip}

\noindent P1QT: Systems

\begin{quote}
Same as P1QM (Section \ref{sotsez:3.2.2}).
\end{quote}

\noindent Definition: Quantal Phase-domain

\begin{quote}
Let $ \mathcal{L} $ denote the real space  of linear, Hermitian
operators on the Hilbert space $ \mathcal{H} $ of the system, and
$ \mathcal{P} $ the convex subset of $ \mathcal{L} $ of all
linear, Hermitian, nonnegative-definite, unit-trace operators $
\rho $. We call $ \mathcal{P} $ the \emph{quantal phase-domain}.
\end{quote}

\noindent P2QT: Homogeneous Preparations

\begin{quote}
To every \emph{homogeneous preparation}  scheme $ \Pi $ for a
system, there corresponds an operator $ \rho $ in the quantal
phase-domain $ \mathcal{P} $ of the system. We call $ \rho $ the
\emph{state operator}.
\end{quote}

\noindent P3QT: Physical Observables

\begin{quote}
Some real functionals $ g, h, \, ... $  defined on $ \mathcal{P} $
correspond to \emph{physical observables} of the system. Given an
ensemble of identical systems prepared according to the
homogeneous scheme $ \Pi $, with corresponding state operator $
\rho $, the arithmetic \emph{mean value} $ \overline{g} $ of data
yielded by measurements of the observable $ g $ is given by the
value of the functional $ g $ at the point $ \rho $ in the quantal
phase-domain, namely, \bd \overline{g} = g ( \rho ) \, . \ed
\end{quote}

\noindent P4QT: States

\begin{quote}
Every individual system is always in  a \emph{state} described by
some state operator $ \rho $ in the corresponding quantal
phase-domain.
\end{quote}

\noindent P5QT: Causal Evolution

\begin{quote}
For every physical system there exists  a superoperator $ \hat{N}
$ (the evolution superoperator), defined on the Hilbert space of
the system, which determines the \emph{causal evolution} of the
state descriptor $ \rho $ via the following law \be \fr{d \rho}{d
t} = \hat{N} ( \rho ) \, . \label{4.1} \ee The expression for the
evolution superoperator $ \hat{N} $ is given in Chapter
\ref{cap:5} (eq. \ref{5.30}). It differs substantially from the
so-called Liouvillian superoperator $ \hat{L}_o $ (eq. \ref{5.3}).
\end{quote}

\noindent It is noteworthy that to every  real, \emph{linear}
functional $ g $ defined on $ \mathcal{L} $ (and therefore also on
$ \mathcal{P} $), there corresponds a unique linear, Hermitian
operator $ G $ on $ \mathcal{H} $ such that $ g ( \rho ) = \Tr ( G
\rho ) $, for every $ \rho $. Thus, the representation of physical
observables by means of linear, Hermitian operators (cf. the
paradigm of Quantum Mechanics) is contained in the present
formulation as a particular case. For example, the physical
observable energy is represented here by the linear functional $ h
( \rho ) = \Tr ( H \rho ) $ where $ H $ is the standard
Hamiltonian operator of the system. However, a new feature of the
proposed paradigm, in addition to the introduction of the
evolution superoperator $ \hat{N} $, is that postulate P3QT has
enlarged substantially the class of physical observable
representatives, to include \emph{nonlinear} functionals on $
\mathcal{P} $. An example of nonlinear   functional  is  $ s (
\rho ) = - k \Tr ( \rho \ln \rho ) $. As   shown   by G.N.
Hatsopoulos and E.P. Gyftopoulos (1976) this nonlinear functional
represents an important physical observable called the
\emph{entropy}.

It is also noteworthy that postulate P4QT,  compared to P4QM, has
enlarged substantially the class of mathematical state
descriptors. Whereas Quantum Mechanics includes only the states
that can be described by idempotent state operators (i.e.,
projection operators $ P_{\psi} $ onto the one-dimensional
subspaces of $ \mathcal{H} $ spanned by the state vectors $ \psi
$), Quantum Thermodynamics includes states that are described by
both idempotent and non-idempotent state operators. This
enlargement of the class of state descriptors has allowed
Hatsopoulos and Gyftopoulos to encompass within a single quantum
theoretical structure both Mechanics and Thermodynamics.

An essential distinction must be noticed  between the paradigm of
Quantum Thermodynamics and that of Quantum Statistical Mechanics,
specifically, the distinction between the \emph{state operator} $
\rho $ and the \emph{statistical} (or density) \emph{operator} $ W
$ (as defined in Section \ref{sotsez:3.2.2}). Though these two
operators have the same \emph{mathematical} properties, they
describe different concepts. A state operator $ \rho $ describes a
\emph{homogeneous} preparation   scheme, while   a   statistical
operator $ W $ describes any preparation scheme, regardless of its
homogeneity. Indeed, Quantum Thermodynamics is a \emph{physical}
theory, while Quantum Statistical Mechanics is a
\emph{statistical} theory. Thus, the physical meaning attached to
these two mathematically similar operators is completely
different. However, within the present formulation, no confusion
should arise, since no statistical or density operator is defined.

To complete the formulation of our general  quantum theory, we
need to superimpose a statistical theory to the paradigm of
Quantum Thermodynamics.

Typically, a statistical theory deals with  situations in which
the preparation scheme is not homogeneous and, thus, there exists
\emph{uncertainty} about which homogeneous preparation scheme is
actually adopted to prepare the individual systems. Uncertanities
of this type are \emph{not} inherent in the nature of the
individual states, rather, they are inherent in the nature of the
preparation scheme and they are linked ono-to-one to the
heterogeneity of the scheme. The purpose of a statistical theory
is to describe all preparation schemes and distinguish
unambiguously between two types of uncertainties: quantal
uncertainties associated with the quantal nature of the states of
each individual system and nonquantal uncertainties associated
with the heterogeneity of a preparation scheme. The next Section
sharpens this distinction.

\subsection{Quantal vs Nonquantal Uncertainties} \label{sotsez:4.1.2}

Consider a heterogeneous preparation  $ \Pi $ resulting from the
statistical composition of different homogeneous schemes $ \Pi_n $
with corresponding statistical weights $ w_n $. An
\emph{individual} system prepared according to such a preparation
scheme is ``actually" prepared according to the homogeneous scheme
$ \Pi_n $ with probability $ w_n $. From the postulates P2QT and
P4QT of Quantum Thermodynamics, it follows that an individual
system prepared according to the homogeneous scheme $ \Pi_n $ is
``in" the \emph{state} described by the corresponding state
operator $ \rho_n $. Therefore, an individual system prepared
according to $ \Pi $ is in \emph{state} $ \rho_n $ with
probability $ w_n $.

The probabilities $ w_n $ are characteristics  of the
heterogeneous preparation $ \Pi $, having as empirical
correspondents the relative populations, in a heterogeneous
ensemble prepared according to $ \Pi $, of systems whose
individual state is $ \rho_n $. It is clear that such
probabilities are \emph{not} inherent characteristics of the
\emph{state} of the individual member systems of the heterogeneous
ensemble. Rather, they are characteristics of the
\emph{collection} of such individual systems, i.e. of the
statistical structure of the ensemble considered as a whole. In
short, the probabilities $ w_n $ reflect \emph{uncertainties} that
are inherent in the \emph{heterogeneity} of the preparation scheme
$ \Pi $ and are completely \emph{independent} of the nature of the
component \emph{homogeneous} preparations $ \rho_n $.

The following natural question arises: Is  it possible to describe
within a single formulation both nonquantal uncertainties of this
type (not contemplated in Quantum Thermodynamics) and quantal
uncertainties as implied by Quantum Thermodynamics, and yet
maintain unambiguously reflected their essential distinction? In
Section \ref{sotsez:3.2.3}, we have answered to part of this
question, by stating three necessary conditions that such a
formulation must satisfy. These conditions have guided us in the
formulation of the statistical theory proposed in the next
Section, which constitutes an affirmative answer to the question.

\section{QUANTUM STATISTICAL THERMODYNAMICS} \label{sez:4.2}

Quantum Statistical Thermodynamics is the  statistical part of the
general quantum theory that we propose. It is concerned with the
description of all preparations for a physical system or,
equivalently, of distributions of states in any ensemble of
systems each of which individually obeys the laws of Quantum
Thermodynamics. It is based on a mathematical description of
preparations which accounts unambiguously for the two types of
uncertainties that are present in an ensemble generated by a
heterogeneous preparation, namely, quantal and nonquantal
uncertainties.

A statistical theory with these  characteristics is presently
lacking and, as we have seen, this fact is a source of conceptual
difficulties. The formulation that we propose is based on a new
measure-theoretic description of preparations.

\subsection{Measure-theoretic Description of Preparations} \label{sotsez:4.2.1}

This Section presents a mathematical  description of preparations
which satisfies the three necessary conditions stated in Section
\ref{sotsez:3.2.3}. The mathematical notion of \emph{measure}
defined on a set (cf., e.g., G. Fano (1971)) is the key notion on
which the proposed description is based.

We recall that every point of the quantal  phase-domain of a
system is a state operator which, according to postulate P4QT of
Quantum Thermodynamics, describes one of the possible states of an
individual system or, equivalently, a homogeneous preparation for
such a system.

We now postulate that the mathematical  descriptor   of a
preparation scheme is a \emph{measure} $ \mu $ defined on the
quantal phase-domain $ \mathcal{P} $ of the system, satisfying the
following normalization condition

\be \mu ( \mathcal{P} ) = \int_{\mathcal{P}} \mu ( d \rho ) = 1 \,
. \label{4.2} \ee

\noindent We call $ \mu $ the  \emph{statistical-weight measure}.
In addition, we postulate that the expected value of a physical
observable represented by the functional $ g $, defined on $
\mathcal{P} $, is given by

\be \langle  \overline{g} \rangle  =  \int_{\mathcal{P}}  g ( \rho
) \mu ( d \rho ) \, . \label{4.3} \ee

We now examine the proposed description  in the light of the
conditions of Section \ref{sotsez:3.2.3}. Condition (i) is
satisfied by defining the rule of measure combination

\be \mu = \sum_n w_n \mu_n \label{4.4} \ee

\noindent corresponding to the concept of  statistical composition
of different preparations. In fact, the corresponding expected
mean value of a generic observable, with corresponding functional
$ g $, is correctly given by

\bd \langle  \overline{g} \rangle  = \int_{\mathcal{P}}  g ( \rho
) \mu ( d \rho ) = \int_{\mathcal{P}} g ( \rho ) \sum_n w_n \mu_n
( d \rho ) = \ed \be = \sum_n w_n \int_{\mathcal{P}} g ( \rho )
\mu_n ( d \rho ) = \sum_n w_n { \langle  \overline{g} \rangle  }_n
\, . \label{4.5} \ee

Among the measures that can be defined on  $ \mathcal{P} $ are the
\emph{Dirac measures} defined as follows. Let $ \rho_o $ be a
state operator and let $ E $ be any subset of $ \mathcal{P} $. The
corresponding Dirac measure $ {\mu_{\rho}}_o $   is (cf. G. Fano
(1971), p.207)

\be {\mu_{\rho}}_o ( E ) = \lgr \begin{array}{ll}  1 & \mbox{if }
\rho_o \in E \\ 0 & \mbox{if } \rho_o \notin E \end{array} \rbl \,
. \label{4.6} \ee

\noindent The Dirac measures are \emph{irreducible}  into weighted
sums of \emph{different} measures, since their \emph{support}
(intended here as that part of the domain $ \mathcal{P} $ for
which the measure is nonzero) is indivisible, being a \emph{single
point} in the phase-domain: the state operator $ \rho_o $.

Since to every homogeneous preparation (i.e.,  to every state
operator $ \rho $) there corresponds a Dirac measure, and the
Dirac measures are irreducible, it follows that condition (ii) of
Section \ref{sotsez:3.2.3} is satisfied.

Condition (iii) is also satisfied.  In fact, the following theorem
holds:

\begin{quote}
every normalized measure $ \mu $ defined on  the domain $
\mathcal{P} $ can be \emph{uniquely} decomposed into a ``weighted
sum" of Dirac measures.
\end{quote}

To prove this theorem we write the generic  measure $ \mu $ as a
``weighted sum" of Dirac measures

\be \mu = \int_{\mathcal{P}} \mu_{\rho}\, \sigma ( d \rho ) \, .
\label{4.7} \ee

\noindent where $ \sigma $ is some measure  expressing the weight
with which each component Dirac measure $ \mu_{\rho} $
participates to form the measure $ \mu $. From the definition of
Dirac measures, it is easy to see that $ \sigma = \mu $, a
\emph{unique} solution. In fact, for any subset $ E $ of $
\mathcal{P} $, equation \ref{4.7} becomes

\be \mu ( E ) = \int_E \sigma ( d \rho ) = \sigma ( E ) \, .
\label{4.8} \ee

\noindent In other words, the generic measure  $ \mu $ represents
the weight with which each Dirac measure participates to form the
measure $ \mu $ itself. This theorem also implies that the Dirac
measures are the only irreducible measures definable over $
\mathcal{P} $. In fact, any other measure can be decomposed into a
``sum" of Dirac measures and is therefore reducible. The physical
meaning of this theorem is that the descriptor of a generic
preparation, i.e. the generic measure $ \mu $, admits of a
\emph{unique ``spectral" resolution} into the descriptors of its
homogeneous component preparations, i.e. the Dirac measures $
\mu_{\rho} $.

We conclude that the mathematical description of  a heterogeneous
preparation by means of a measure $ \mu $ defined on the quantal
phase-domain is adeguate, unambiguous and does not lead to
paradoxes. We are finally in the position to propose the paradigm
of Quantum Statistical Thermodynamics.

\subsection{The Paradigm of Quantum Statistical Thermodynamics} \label{sotsez:4.2.2}

The paradigm of Quantum Statistical Thermodynamics  is based on
the following five postulates.

\vspace{\baselineskip}

\noindent P1QST: Systems

\begin{quote}
Same as P1QT (Section \ref{sotsez:4.1.1}).
\end{quote}

\noindent Definition: Quantal Phase-domain

\begin{quote}
Same as in Section \ref{sotsez:4.1.1}.
\end{quote}

\noindent P2QST: Preparations

\begin{quote}
To every \emph{preparation} scheme $ \Pi $ for a system, there
corresponds a \emph{measure} $ \mu $ defined on the quantal
phase-domain $ \mathcal{P} $ of the system, satisfying the
normalization condition \bd \qquad \qquad \mu ( \mathcal{P} ) =
\int_{\mathcal{P}} \mu ( d \rho ) = 1 \, . \qquad \qquad \qquad
\qquad (\ref{4.2}) \ed We call $ \mu $ the
\emph{statistical-weight  measure}.
\end{quote}

\break

\noindent P3QST: Physical Observables

\begin{quote}
Some real functionals $ g, h, \, ... $ defined  on $ \mathcal{P} $
correspond to \emph{physical observables} of the system. Given an
ensemble of identical systems prepared according to the scheme $
\Pi $, with associated statistical-weight measure  $ \mu $, the
\emph{expected mean value} $ \langle  \overline{g} \rangle  $ of
measurements of the observable $ g $ is given by \bd \qquad \qquad
\langle  \overline{g} \rangle  = \int_{\mathcal{P}} g ( \rho ) \mu
( d \rho ) \, . \qquad \qquad \qquad \qquad \quad (\ref{4.3}) \ed
\end{quote}

\noindent P4QST: States

\begin{quote}
Same as P4QT. I.e., every system is always  in a \emph{state}
described by some \emph{state operator} $ \rho $ in the
corresponding quantal phase-domain.
\end{quote}

\noindent P5QST: Causal Evolution

\begin{quote}
Same    as  P5QT.   The corresponding   evolution of    the
statistical-weight measure is given by the equation \be \fr{d
\mu}{d t} = \int_{\mathcal{P}} \fr{d \mu_{\rho}}{d t}\, \mu ( d
\rho ) \label{4.9} \ee where \be \fr{d \mu_{\rho}}{d t} = \lim_{dt
\tend 0} \fr{\mu_{\rho} + \hat{N} ( \rho ) dt - \mu_{\rho} }{d t}
\label{4.10} \ee expresses the evolution of the Dirac measures
under the motion generated by the superoperator $ \hat{N} $.
\end{quote}

\vspace{\baselineskip}

\noindent The solution of the limit defining  the evolution of a
Dirac measure (eq. \ref{4.10}) in terms of the evolution
superoperator $ \hat{N} $ as given in Chapter \ref{cap:5} (eq.
\ref{5.30}) is an interesting unsolved problem. Another
interesting open problem is a complete mathematical
characterization of the measure-theoretic properties of the
quantal phase-domain. For example, it would be important to
establish whether a Lebesgue measure can be defined over the
quantal phase-domain. This measure might be the descriptor
associated with the ``most" heterogeneous preparation, obtained by
uniform composition of all the possible homogeneous preparations
for a system.

The following theorems of Quantum Statistical  Thermodynamics
summarize the results of the last Section, where we already
concluded that the description of preparations by means of a
statistical-weight measure is unambiguous.

\vspace{\baselineskip}

\noindent Th1QST

\begin{quote}
Among the normalized measures definable on  the phase-domain $
\mathcal{P} $, only the Dirac measures $ \mu_{\rho} $ are
\emph{irreducible}, in the sense that the equality \be \mu_{\rho}
= w_1 \mu_1 + w_2 \mu_2 \quad \mbox{with} \quad w_1 , \, w_2 > 0
\label{4.11} \ee holds if, and only if, $ \mu_1 = \mu_2 =
\mu_{\rho} $.
\end{quote}

\break

\noindent Th2QST

\begin{quote}
To every distinct state operator $ \rho $  in $ \mathcal{P} $
there corresponds a distinct Dirac measure $ \mu_{\rho} $ defined
on $ \mathcal{P} $.
\end{quote}

\noindent Definition: Homogeneous preparations within QST

\begin{quote}
A preparation is said to be homogeneous  if, and only if, it is
described by a Dirac measure.
\end{quote}

\noindent Th3QST

\begin{quote}
Among all the preparation schemes for a  physical system, only the
\emph{homogeneous} preparations cannot be conceived as the result
of a statistical composition of \emph{different} preparation
schemes. Moreover, each homogeneous preparation corresponds to one
and only one \emph{state} of an individual system, and vice versa.
\end{quote}

\noindent Th4QST

\begin{quote}
Every normalized measure $ \mu $ defined on $ \mathcal{P} $ can be
uniquely expressed as a ``weighted sum" of Dirac measures, i.e.
\bd \qquad \qquad \mu = \int_{\mathcal{P}} \mu_{\rho}\, \sigma ( d
\rho ) \, . \qquad \qquad \qquad \qquad \qquad (\ref{4.7}) \ed
where the measure $ \mu $ itself represents  the ``weight" of the
corresponding Dirac measure. Consequently, every heterogeneous
preparation can be \emph{uniquely resolved} into its homogeneous
component preparations.
\end{quote}

\subsection{Empirical Determination of the Statistical-weight Measure} \label{sotsez:4.2.3}

A measure $ \mu $ defined on the quantal  phase-domain $
\mathcal{P} $ can be ``represented" as follows.  Let $ \mathcal{L}
$ be the real space of linear, Hermitian operators on the Hilbert
space of the system. Let the scalar product $ (\ , \ ) $ in $
\mathcal{L} $ be defined by the trace formula

\be ( F ,  G ) = \Tr ( FG ) \, . \label{4.12} \ee

\noindent Two operators in $ \mathcal{L} $  are said to be
orthogonal if their scalar product vanishes.  Let the operators $
Q_i $ and $   R_j $ form two bases for $ \mathcal{L} $ satisfying
the following orthonormalization conditions:

\be \Tr ( Q_i Q_n ) = \delta_{in} \qquad \Tr  ( R_j R_m ) =
\delta_{jm} \, . \label{4.13} \ee

\noindent The state operators are elements  of the domain $
\mathcal{P} $ which is a convex subset of $ \mathcal{L} $ and may
therefore be expanded in terms of such bases:

\begin{subequations}
\be \rho = \sum_i q_i Q_i = \sum_j r_j R_j \label{4.14a} \ee

\noindent where

\be q_i = \Tr ( \rho Q_i ) \qquad r_j = \Tr ( \rho R_j ) \, .
\label{4.14b} \ee
\end{subequations}

\noindent As discussed by W. Band and  J.L. Park (1977a), once a
basis is specified, the ``coordinates" $ q_i $ (or $ r_j $) of any
state operator identify a point in an auxiliary real space. As the
state operator sweeps the quantal phase-domain, the point in the
auxiliary space sweeps a convex region $ \mathcal{D}_Q $ (or $
\mathcal{D}_R $).

Once the domain $ \mathcal{P} $ has been  coordinatized, any
functional $ g ( \rho ) $ defined on it can be written as a
function of the corresponding coordinates $ g_Q ( \sot{q} ) $ (or
$ g_R ( \sot{r} ) $) defined on the corresponding region of the
auxiliary real space. Moreover, any measure $ \mu $ defined on $
\mathcal{P} $ can generally be written as follows

\be \mu ( d \rho ) = w_Q ( \sot{q} ) d  \sot{q} = w_R ( \sot{r} )
d \sot{r} \label{4.15} \ee

\noindent where the functions $ w_Q $ or  $ w_R $ ``represent" the
measure with respect to the corresponding coordinatization of the
phase-domain. Accordingly, the expected mean value of a physical
observable represented by the functional $ g $ is given by

\be \langle  \overline{g} \rangle  = \int_{\mathcal{P}}  g ( \rho
) \mu ( d \rho ) = \int_{\mathcal{D_Q}} g_Q ( \sot{q} ) w_Q (
\sot{q} ) d \sot{q} = \int_{\mathcal{D_R}} g_R ( \sot{r} ) w_R (
\sot{r} ) d \sot{r} \, . \label{4.16} \ee

In view of this background, we can now  pose the following
question: Is it possible to identify a list, or ``quorum", of
independent physical observables such that the empirical
determination of their expected mean values is tantamount to the
empirical determination of the statistical-weight measure $ \mu $?
This question is composed of two equally important subproblems.
The first subproblem is mathematical and it is to determine
whether there exists a set of functionals $ g_i ( \rho ) $ (not
necessarily linear) on $ \mathcal{P} $ such that the system of
equations

\be \int_{\mathcal{P}} g_i ( \rho ) \mu  ( d \rho ) = \langle
{\overline{g}}_i \rangle  \label{4.17} \ee

\noindent can be ``solved" for the unknown  $ \mu $, once all the
$ \langle  {\overline{g}}_i \rangle  $'s are given. The second
subproblem is to identify the physical meaning of such
functionals.

We have not solved these two problems.  However, for the first
subproblem we can suggest two interesting ``moment" generating
functions which might prove to be sufficient for the empirical
determination of the statistical-weight measure. The first
generating function is

\be f ( u ) = \int_{\mathcal{P}} \Tr ( \exp  ( u \rho ) ) \mu ( d
\rho ) \label{4.18} \ee

\noindent where $ u $ may be a complex  variable. If the function
$ f ( u ) $ had been empirically determined, then the following
sequence of mean values would be known

\be { \lbl \fr{\partial^n f}{\partial u^n}  \rat }_{u = 0} =
\int_{\mathcal{P}} \Tr ( \rho^n ) \mu ( d \rho ) \, . \label{4.19}
\ee

\noindent The second generating function is

\be g' ( u ) = \int_{\mathcal{P}} \exp  ( u g ( \rho )) \mu ( d
\rho ) \label{4.20} \ee

\noindent where $ g $ may be any functional  on $ \mathcal{P} $.
If $ g' ( u ) $ were known, then the following sequence of mean
values would be known

\be { \lbl \fr{\partial^n g'}{\partial u^n}  \rat }_{u = 0} =
\int_{\mathcal{P}} g^n ( \rho ) \mu ( d \rho ) \, . \label{4.21}
\ee

It remains to be proved that the empirical  determination of one
of these generating functions or of a sequence of them is
mathematically equivalent to the empirical determination of the
measure   $ \mu $.

\subsection{Quantum Statistical Thermodynamics and Information Theory} \label{sotsez:4.2.4}

In this Section, we consider three problems  pertaining to the
domain of Information Theory. Here, these problems acquire
physical significance within Quantum Statistical Thermodynamics.
However, their conceptual nature is independent of the formulation
of QST and it has been widely discussed in the literature, with or
without reference to other statistical theories (e.g., cf. R.D.
Levine and M. Tribus (1978) and, for a recent mathematical
treatment, N.F.G. Martin and J.W. England (1981); it is to be
noticed that, in this literature, the terms ``informational
entropy" or, simply, ``entropy" refer to what we shall call
\emph{statistical uncertainty}, in order to avoid confusion
between those information-theoretic concepts and the quantum
thermodynamical observable entropy of any individual physical
system).

The three problems are about an experimenter  or ``observer"
accepting for study \emph{individual} systems prepared according
to a heterogeneous preparation scheme. They are:

\begin{quote}\begin{description}
    \item[(a) ] The observer knows the statistical-weight measure
corresponding to the preparation (e.g., he has determined it by
previous experiments). Therefore, he knows completely the
statistical structure of the heterogeneous preparation, since, by
analysis of the statistical-weight measure, he can identify the
component homogeneous preparations and their statistical weights.
However, he cannot predict according to which of the component
homogeneous schemes the next single system is going to be actually
prepared. The first information-theoretic problem is to define an
\emph{indicator} of the uncertainty of the observer about the
outcome of the next single act of preparation. In other words,
this indicator should globally quantify the heterogeneity of the
preparation scheme. This problem is conceptually identical to that
considered by C.E. Shannon (1948) in his famous article.
    \item[(b) ] The observer does not know the statistical-weight
    measure of the preparation, however, he knows which are its
    component homogeneous preparations (i.e., by some means he has
    determined the ``spectrum" of the statistical-weight measure).
    In addition, he also knows the expected values of a limited
    number of physical observables. The second
    information-theoretic problem is to find the most probable
    statistical-weight measure consistent with the given expected
    mean values and the additional available information about the
    component homogeneous preparations.
    \item[(c) ] The observer does not  know the statistical-weight
    measure of the preparation. All he knows are the expected mean
    values of a limited number of physical observables. The third
    information-theoretic problem is also to find the most
    probable statistical-weight measure consistent with the given
    expected mean values. Differently from the previous problem,
    here the spectrum of the statistical-weight measure is also an
    unknown. The last two problem are conceptually identical to
    those considered by E.T. Jaynes (1957) in his famous and
    subsequent articles.
\end{description}\end{quote}

Problem (a) can be solved by defining an  indicator $ I_{\mu} $ of
the heterogeneity of a preparation scheme described by the
statistical-weight measure $ \mu $. The    indicator   $ I_{\mu} $
must globally quantify an objective characteristic -- the
heterogeneity -- of the preparation. Since this characteristic is
already contained in the statistical-weight measure, the indicator
should be defined in terms of the properties of such measure only.
If the resolution of $ \mu $ into Dirac measures identifies a
countable spectrum, i.e.

\bd \qquad \qquad\qquad \qquad\qquad \mu = \sum_n w_n  {
\mu_{\rho} }_n \, , \qquad \qquad \qquad\qquad\qquad (\ref{4.4})
\ed

\noindent then, by analogy with Shannon's  problem (cf. eq.
\ref{2.1}), we can define

\begin{subequations}
\be I_{\mu} = - c \sum_n w_n \ln w_n \, . \label{4.22a} \ee

\noindent It is noteworthy that $ 0  \leq I_{\mu} \leq c \ln N $,
where $ N $ is the number of points forming the spectrum of the
measure and may be finite or infinite. If the spectrum is instead
not countable, then it is possible to define (as we will see for
problem (c)) indicators that are relative to some \emph{a priori}
partition of the quantal phase-domain. However, such definitions
are not suitable to solve the present problem, because a priori
partitions are not inherent characteristics of the measure $ \mu
$. Instead, if the spectrum is not countable, we set

\be I_{\mu} = \infty \label{4.22b} \ee
\end{subequations}

\noindent and justify this definition  arguing that in this case
the heterogeneity is greater than any of the cases with countable
resolution, thus including those with $ I_{\mu} = c \ln N $ and $
N = \infty $.

We call the indicator   $ I_{\mu} $ the  \emph{statistical
uncertainty} associated with the preparation scheme described by $
\mu $. This definition is of key importance to approach problems
(b) and (c). In fact, essential to their solution is Jaynes'
\emph{maximum statistical uncertainty principle}. As we have seen
in Chapter \ref{cap:2}, this principle is known as the
\emph{maximum ``entropy" principle}, however, the term entropy has
already been used in Quantum Thermodynamics to indicate an
important physical observable of any material system. Thus, we
shall not use it to indicate also the present
information-theoretic concept. This principle, in our case,
prescribes that, among the statistical-weight measures sharing the
known expected mean values and the remaining known
characteristics, the most probable measure is that with the
maximum value of the indicator of statistical uncertainty, i.e.
the most heterogeneous.

Problem (b) can be readily solved by means  of Jaynes' principle
when the known set of component homogeneous preparations is
countable (i.e., the spectrum of the statistical-weight measure is
countable). Assume, for example, that the only known expected mean
value is that of the energy. Then, the most probable
statistical-weight measure is to be found by maximizing

\bd I_{\mu} = - c \sum_n w_n \ln w_n \ed

\noindent subject to the constraint

\be \langle  \overline{h} \rangle  = \int_{\mathcal{P}}  h ( \rho
) \mu ( d \rho ) = \sum_n w_n { \langle  \overline{h} \rangle  }_n
\label{4.23} \ee

\noindent where $ \langle  \overline{h} \rangle  $ is  known
(presumably, as a result of previous experiments) and

\be { \langle  \overline{h} \rangle  }_n = \Tr ( H \rho_n )
\label{4.24} \ee

\noindent are also known, since the  spectrum of the measure is
known and so are therefore the corresponding state operators $
\rho_n $. Maximization yields the following most probable
statistical-weight measure

\be \mu = \sum_n q_n { \mu_{\rho} }_n \label{4.25} \ee

\noindent where

\be q_n = \fr{\exp ( - b { \langle  \overline{h} \rangle   }_n
)}{\sum_m \exp ( - b { \langle  \overline{h} \rangle  }_m )}
\label{4.26} \ee

\noindent and the parameter $ b $ is fixed  by the given expected
mean value of the energy. When the spectrum of the
statistical-weight measure is known but not countable, the
solution involves difficulties as discussed below for problem (c).

Problem (c) can also be given a solution  which is not, however,
fully satisfactory. This solution is based on a different
definition of statistical uncertainty for the case in which the
spectrum of the statistical-weight measure is, a priori, the whole
quantal phase-domain (or some known region of it, for problem
(b)). An important, and somewhat unsatisfactory, feature of the
following definition is that it depends not only on the
statistical-weight measure but also on an \emph{a priori} selected
countable partition of the quantal phase-domain. Let the
collection of disjoint subsets $ E_n $ form a particular countable
partition $ \zeta$ of the quantal phase-domain $ \mathcal{P} $.
With respect to this partition, the following definition of
statistical uncertainty can be given

\be I_{\mu} ( \zeta ) = - c \sum_n \mu  ( E_n ) \ln \mu ( E_n ) \,
. \label{4.27} \ee

\noindent Maximization of this indicator,  subject to the
constraints expressing the known expected mean values, yields the
most probable statistical-weight measure relative to partition $
\zeta $. Again, if the only expected mean value is the energy, we
write

\be \langle  \overline{h} \rangle  = \int_{\mathcal{P}}  h ( \rho
) \mu ( d \rho ) = \sum_n \int_{E_n} h ( \rho ) \mu ( d \rho ) =
\sum_n \mu ( E_n ) \langle  \overline{h} ( E_n ) \rangle
\label{4.28} \ee

\noindent where the last    equality     defines the expected mean
energy $ \langle  \overline{h} ( E_n ) \rangle  $ within the $ n
$-th cell of the partition. Maximization in this case yields

\be \mu ( E_n ) = \fr{\exp ( - b \langle   \overline{h} ( E_n )
\rangle )}{\sum_m \exp ( - b \langle  \overline{h} ( E_m ) \rangle
)} \, . \label{4.29} \ee

\noindent This expression does not  identify a single most
probable statistical-weight measure but a family of measures
sharing these values on the subsets $ E_n $. A more unsatisfactory
aspect of this solution is its dependence on the choice of the a
priori partition of phase-domain. Different a priori partitions
would yield different solutions. Since, by the definition of the
problem, no information is available about the structure of the
preparation and therefore on the statistical-weight measure, we
need an additional criterion (perhaps based on the nature of the
constraints or on the type of estimates that one plans to make
with the result) to choose among different partitions of quantal
phase-domain. Presumably, this as yet unknown criterion is to be
postulated independently of, and in addition to, the maximum
statistical uncertainty principle.

The following example sharpens the  distinction between the
physical observable entropy and the information-theoretic
indicator of statistical uncertainty.

\subsection{Example. Entropy vs Statistical Uncertainty} \label{sotsez:4.2.5}

Let $ \Pi_1 $ and $ \Pi_2 $ be two  distinct homogeneous
preparations for a given system. We want to describe the
heterogeneous preparation corresponding to the symbolic expression
(cf. Section \ref{sotsez:3.1.2})

\bd \Pi = w_1 \Pi_1 + w_2 \Pi_2 \qquad  \qquad \mbox{with} \quad
w_1 + w_2 = 1 \mbox{ and } w_1 , \, w_2 > 0 \, . \ed

\noindent According to the paradigm of  Quantum Thermodynamics,
the two homogeneous preparations are described by the state
operators $ \rho_1 $ and $ \rho_2 $, respectively. We now define
the measure $ \mu $ corresponding to the heterogeneous preparation
$ \Pi $. Let $ E $ be any subset of the quantal phase-domain $
\mathcal{P} $ of the system. Then

\be \mu ( E ) = \lgr \begin{array}{ccccc}  w_1 & \mbox{if} &
\rho_1 \in E & \mbox{and} & \rho_2 \notin E \\ w_2 & \mbox{if} &
\rho_1 \notin E & \mbox{and} & \rho_2 \in E \\ 1 & \mbox{if} &
\rho_1 \in E & \mbox{and} & \rho_2 \in E \\ 0 & \mbox{if} & \rho_1
\notin E & \mbox{and} & \rho_2 \notin E \end{array} \rbl
\label{4.30} \, . \ee

\noindent It is easy to check that $ \mu $  is a measure defined
on $ \mathcal{P} $ and that the normalization condition $ \mu (
\mathcal{P} ) = 1 $ is satisfied.

From the definition of $ \mu $, it is  immediately seen that

\be \mu = w_1 { \mu_{\rho} }_1 + w_2 { \mu_{\rho} }_2 \, .
\label{4.31} \ee

\noindent This represents the spectral  resolution of $ \mu $ into
its component Dirac measures. Interestingly, the expected mean
value of the physical observable \emph{entropy}, represented by
the functional $ s ( \rho ) = - k \Tr ( \rho \ln \rho ) $, is
given by

\bd \langle  \overline{s} \rangle  = - k \int_{\mathcal{P}} \Tr (
\rho \ln \rho ) \mu ( d \rho ) = \ed \be = - k w_1 \Tr ( \rho_1
\ln \rho_1 ) - k w_2 \Tr ( \rho_2 \ln \rho_2 ) = w_1
{\overline{s}}_1 + w_2 {\overline{s}}_2 \, . \label{4.32} \ee

\noindent It is important to note that this  expression is
unequivocally different from the value of the indicator of
\emph{statistical uncertainty} for the same heterogeneous
preparation. In fact, the corresponding expression for $ I_{\mu} $
is

\be I_{\mu} = - c ( w_1 \ln w_1 + w_2 \ln w_2 ) \, . \label{4.33}
\ee

\noindent As explained above, this quantity is an  index of the
heterogeneity of the preparation scheme and is totally independent
of the physical nature of the states of systems prepared according
to such scheme, i.e. of the nature of the component homogeneous
preparations. In fact, equation \ref{4.33} is totally independent
of the state operators $ \rho_1 $ and $ \rho_2 $.

\chapter{NEW GENERAL QUANTUM THEORY: DYNAMICS} \label{cap:5}

This Chapter completes the dynamical postulate P5QT of Quantum
Thermodynamics by proposing a general expression for the evolution
superoperator $ \hat{N} $. The new law of causal evolution has
been conceived so as to entail both a statement of the second law
of Thermodynamics and a statement of existence of irreversible
processes. Conversely, these two statements emerge from Quantum
Thermodynamics as manifestations of the fundamental quantum
dynamical behaviour of the elementary constituents of any
individual material system.

An introductory discussion and some necessary preliminary
definitions are presented in Section \ref{sez:5.1}. The equation
of motion is proposed in Section \ref{sez:5.2}.

\section{GENERAL REMARKS AND DEFINITIONS} \label{sez:5.1}

According to postulate P5QT, the general law of causal evolution
of Quantum Thermodynamics has of the form

\bd \qquad \qquad \qquad \fr{d \rho}{d t} = \hat{N} ( \rho )
\qquad \qquad \qquad (\ref{4.1}) \ed

\noindent where $ \hat{N} $ is the evolution superoperator of the
system and $ \rho $ is the state operator. Before introducing the
explicit form of the evolution superoperator $ \hat{N} $, we
examine the motivations for developing a new nonlinear
irreversible quantum dynamics.

\subsection{Incompleteness of Hamiltonian Dynamics} \label{sotsez:5.1.1}

As shown by F. Strocchi (1966), the Schr\"odinger equation of
motion (eq. \ref{3.8}) can be viewed as a natural extension to
Quantum Mechanics of the Hamilton equations of motion of Classical
Mechanics (eq. \ref{3.1}). As we have noticed in Section
\ref{sotsez:4.1.1}, in Quantum Thermodynamics the states of
Quantum Mechanics are described by a particular subset of state
operators, namely, the projection operators $ P_{\psi} $ onto
one-dimensional subspaces of the Hilbert space spanned by the
vectors $ \psi $ (the state vectors of Quantum Mechanics). As the
state vector evolves according to the Schr\"odinger equation, the
corresponding projection operator evolves according to the
equation

\be \fr{d P_{\psi}}{d t} = - \, \fr{i}{\hbar} [ H ,  P_{\psi} ]
\label{5.1} \ee

\noindent where $ H $ is the Hamiltonian operator of the system.
The extension of equation \ref{5.1} to describe the evolution of
any state operator of Quantum Thermodynamics, i.e.

\be \fr{d \rho}{d t} = - \, \fr{i}{\hbar} [ H  ,  \rho ] \, ,
\label{5.2} \ee

\noindent will be said to describe the Hamiltonian evolution of
the state operator. The corresponding evolution superoperator is
the so-called \emph{Liouvillian superoperator}

\be \hat{L}_o ( \rho ) = - \, \fr{i}{\hbar} [H ,  \rho] \, .
\label{5.3} \ee

\noindent A mathematically identical equation was obtained by J.
von Neumann (1932) for the statistical operator $ W $ of Quantum
Statistical Mechanics (eq. \ref{3.12}). However, the physical
significance of the von Neumann equation is completely different
from equation \ref{5.3}, due to the essential difference between a
state operator $ \rho $ and a statistical operator $ W $ (cf.
Section \ref{sotsez:4.1.1}).

A process of an \emph{isolated system} (the object of study of
Quantum Thermodynamics) is said to be reversible (irreversible) if
the mean value of the entropy during the process remains constant
(increases). The incompleteness of Hamiltonian dynamics regards
both the notion of \emph{irreversibility} and that of
\emph{reversibility}. It stems from the following reasoning:

\begin{quote}\begin{description}
    \item[(a) ] A theorem of the Hamiltonian evolution generated
    by the Liouvillian superoperator is: the mean value of the
    entropy is a constant of the motion for an isolated system.
    According to Quantum Thermodynamics, entropy is a physical
    property of matter. Therefore, if irreversibility exists, then
    it must be a fundamental physical phenomenon since, by
    definition, it must entail an increase in the physical
    property entropy. Common experience shows that most physical
    processes of isolated systems are irreversible.    Therefore,
    entropy cannot be a constant of the motion for isolated
    systems. Hence, Hamiltonian dynamics is incompatible with
    irreversible physical reality.
    \item[(b) ] The entropy functional can be written as follows
    \be s ( \rho ) = - k \Tr ( \rho \ln \rho ) = - k \sum_i p_i \ln
    p_i \label{5.4} \ee where   $ p_i $ is the  $ i $-th
    eigenvalue of the state operator $ \rho $ and the sum runs
    over a complete set of eigenvectors of $ \rho $. A theorem of
    the Hamiltonian evolution generated by the Liouvillian
    superoperator is: each eigenvalue $ p_i $ of the state
    operator is a constant of the motion for an isolated system.
    Such a theorem implies that the entropy remains constant in a
    very special way (Hamiltonian reversibility) and that there
    could not exist reversible processes in which the entropy
    remains constant while the eigenvalues $ p_i $ vary. However,
    as shown by Hatsopoulos and Gyftopoulos (1976), there exist
    examples of important reversible processes of this type.
    Hence, Hamiltonian dynamics is incomplete even as regards the
    notion of reversibility.
\end{description}\end{quote}

These two considerations motivate the search for a more general
equation of motion for Quantum Thermodynamics.

\subsection{Inadequacy of Generalized Linear Dynamics} \label{sotsez:5.1.2}

In the search for a more general equation of motion, a logical
approach is to consider \emph{linear} evolution superoperators
that generalize the Liouvillian superoperator (cf. J.L. Park and
W. Band (1977b)). In appendix \ref{app:B} (eq. \ref{B.2}) we
consider the explicit form of the most general linear evolution
superoperator $ \hat{L} $ that could be postulated to generate the
motion of the state operator via the following equation

\be \fr{d \rho}{d t} = \hat{L} ( \rho ) \, . \label{5.5} \ee

We take the position that such generalized linear equation of
motion is inadequate for the purposes of Quantum Thermodynamics.
The inadequacy regards the notions of natural tendency towards a
stable equilibrium state and of rate of entropy production. It
stems from the following reasoning:

\begin{quote}\begin{description}
    \item[(a) ] It is thermodynamically legitimate, though not
    strictly necessary, to require that the general equation of
    motion eventually transform any initial state of an isolated
    system into the corresponding stable equilibrium state with
    the same mean values for the energy and the other constants of
    the motion. This feature would represent a genuine natural
    tendency towards a stable equilibrium state. However, R.S.
    Simmons (1979) has proved that, for a system whose Hilbert
    space is more than two-dimensional, there exist \emph{no}
    linear equations of motion capable of eventually transforming
    \emph{all} given initial state operators into the
    corresponding stable equilibrium state operators. Hence, the
    notion of natural tendency towards stable equilibrium states
    cannot be described by a linear equation of motion.
    \item[(b)
    ] In appendix \ref{app:B} we give an expression (eq.
    \ref{B.7a}) for the rate of entropy production implied by the
    explicit form of the most general linear evolution
    superoperator $ \hat{L} $. It follows from such expression
    that if the state operator is singular (i.e., at least one of
    its eigenvalues is zero), then the rate of entropy production
    can generally be \emph{infinite}. This practically unknown
    characteristic feature of linear dynamics is not, strictly, in
    conflict with the laws of either Mechanics or Thermodynamics.
    However, it implies that a large class of states (including
    all the states contemplated in Quantum Mechanics) are highly
    unstable and tend towards more stable states at an infinite
    rate.
\end{description}\end{quote}

These considerations led us to consider \emph{nonlinear} evolution
superoperators.

\subsection{Constitutive Structure of a Physical System} \label{sotsez:5.1.3}

We will see that the evolution superoperator $ \hat{N} $ (eq.
\ref{5.30}) is strongly nonlinear in the state operator and its
form depends explicitly on the structure of the system. For
example, the form of $ \hat{N} $ for a single structureless
electron differs substantially from the form of $ \hat{N} $ for a
single electron with internal structure. Thus, according to the
proposed nonlinear dynamics, these systems must be considered as
two \emph{essentially} different physical systems.

Given any physical system, it is necessary to specify exactly its
\emph{constitutive structure} in order to write the corresponding
evolution superoperator. The constitutive structure of a system is
defined here as the list of its \emph{elementary constituents}.

The term \emph{elementary constituent} of a material system is
taken to be a \emph{primitive concept} of the present theory,
analogous to the concept of ``particle" in Newtonian mechanics.
What is judged elementary in physics is of course subject to
sometimes rapid historical evolution;  thus the actual admissible
referents of the term elementary constituent should be left
unspecified in the present general discussion. However, to fix the
ideas, we give the following examples of elementary constituents:
a single fermion (electron, nucleon, ...) or boson (photon, meson,
...), a fermion or boson field (e.g., the electromagnetic field).

To every system composed of a single elementary constituent,
according to postulate P1QT of Quantum Thermodynamics, there
corresponds a Hilbert space. The Hilbert space corresponding to a
general composite system, i.e. a composite of $ M $
distinguishable elementary constituents, is

\be \mathcal{H} = \bigotimes^M_{J = 1} \mathcal{H} ( J ) =
\mathcal{H} ( 1 ) \otimes \mathcal{H} ( 2 ) \otimes \cdots \otimes
\mathcal{H} ( M ) \label{5.6} \ee

\noindent where $ \mathcal{H} ( J ) $ is the ``factor'' Hilbert
space corresponding to the $ J $-th elementary constituent. We
will see that, in the framework of Quantum Thermodynamics, this
``factorization" of the Hilbert space $ \mathcal{H} $ acquires an
important dynamical significance, since the structure of the
evolution superoperator depends explicitly on it.

Mathematically, any given Hilbert space can be factored in many
different ways. Within Hamiltonian dynamics these different
factorizations do not play any role and, therefore, are all
equivalent. However, within the present nonlinear dynamics,
different factorizations correspond to different physical systems,
since they yield different equations of motion. Therefore, it is
essential that the proper selection of the elementary constituents
be made.

\subsection{Generators of the Motion} \label{sotsez:5.1.4}

In conventional quantum dynamics, the term \emph{generator of the
motion} is synonymous with the Hamiltonian. Here, however, we
extend this notion to refer to members of a set of operators which
enter explicitly into the dynamical law and in that sense
determine, or ``generate", the motion.

We postulate that to every type of elementary constituent there
corresponds a list of operators (the \emph{generators of the
motion})

\bd H , \, X , \, ... , \, Y \ed

\noindent defined on the Hilbert space of a system composed of the
elementary constituent only. We further postulate that the first
generator of the list is always the standard Hamiltonian operator
$ H $ associated with the elementary  constituent, and    that the
remaining generators $ X , \, ... , \, Y $ are
\emph{dimensionless} Hermitian operators all commuting with $ H $.
We call them the \emph{non-Hamiltonian generators} of the
elementary constituent. For generality, their number and physical
significance will be left unspecified, since they depend on the
particular type of elementary constituent under consideration.
However, we will see that a theorem of the new dynamics is that
the generators of the motion, together with the identity operator,
form a complete set of constants of the motion for the elementary
constituent. This result should be used to identify the
non-Hamiltonian generators associated with different elementary
constituents. For example, candidates for this role are: the
number of particles operator (if the elementary constituent is a
field), the angular momentum operator, the momentum operator, etc.
Identification of the non-Hamiltonian generators associated with
the known elementary constituents of matter has not been attempted
in this work and is an open avenue for further theoretical and
experimental investigation.

\subsection{Separable Systems and Independent States} \label{sotsez:5.1.5}

We have said that the object of study of Quantum Thermodynamics is
any \emph{isolated} physical system. In fact, the equation of
motion of Quantum Thermodynamics, as proposed in the next
Sections, applies only to isolated systems. Here, we give an
explicit definition of the term isolated system, together with
some preliminary definitions and corollaries of Quantum
Thermodynamics that are independent of the dynamical postulate and
will serve in the definition of the evolution superoperator for a
general composite system.

Let us consider a general composite system as defined in Section
\ref{sotsez:5.1.3}. The index $ J $ identifies the elementary
constituents of the system and runs over the set of labels $ 1 ,
\, 2 , \, ... , \, M $. The Hilbert space of such a system is

\bd \qquad \qquad \qquad \mathcal{H} =\bigotimes^M_{J = 1}
\mathcal{H} ( J ) \, . \qquad \qquad \qquad (\ref{5.6}) \ed

\noindent Let us consider a partition of the composite system into
subsystems. The elementary constituents are grouped into disjoint
subsets forming the subsystems. Correspondingly, the set of labels
$ 1 , \, 2 , \, ... , \, M $ is partitioned into disjoint subsets
(e.g, $ (1,3) , \, (6,5,2) , \, ... , \, (8,M) $) which we index
by the letter $ K $. The index $ J ( K ) $ will identify the $ J
$-th elementary constituent of the $ K $-th subsystem and run over
the set $ 1 , \, 2 , \, ... , \, M ( K ) $,  where $ M ( K ) $ is
the total number of elementary constituents of subsystem $ K $.
The Hilbert space corresponding to subsystem $ K $ is

\be \mathcal{H} ( K ) = \bigotimes^{M ( K )}_{J ( K ) = 1}
\mathcal{H} ( J ( K ) ) \, . \label{5.7} \ee

\noindent An apostrophe following an index shall identify the
``complementary" of what is otherwise identified by the index. For
example, $ J' $ identifies the subsystem of the composite system
formed by all the elementary constituents except the $ J $-th one;
$ K' $ the subsystem of the composite system formed by all the
subsystems except the $ K $-th one; $ J'( K ) $ all the elementary
constituents of subsystem $ K $ except the $ J( K ) $-th one; $ J
( K )' $ the subsystem of the composite system formed by all the
elementary constituents except the $ J ( K ) $-th one; finally, $
J ( K' ) $ the $ J $-th elementary constituent of subsystem $ K'
$.

\vspace{\baselineskip}

\noindent Definition: Separable Subsystems

\begin{quote}
A subsystem $ K $ of a composite system is said to be
\emph{separable} if, and only if, the Hamiltonian operator of the
composite system can be written as \be H = H ( K ) \otimes I ( K '
) + I ( K ) \otimes H ( K ' ) \label{5.8} \ee where $ H ( K  ) $
is the Hamiltonian operator on $ \mathcal{H} ( K ) $ corresponding
to a system composed of subsystem $ K $ only. The operator $ H ( K
) $ can be called the \emph{private} Hamiltonian operator of the $
K $-th subsystem.
\end{quote}

\noindent Definition: Separable Systems

\begin{quote}
A system is said to be \emph{separable} if, and only if, it is a
separable subsystem of every conceivable composite system which
contains it.
\end{quote}

\noindent Corollary: Additivity of Energy for Separable Subsystems

\begin{quote}
The mean value of the energy for a system composed of two
\emph{separable} subsystems $ K $ and $ K' $ is equal to the sum
of the ``private energy" values for two subsystems, i.e. \bd
\overline{h} = {\Tr}_{\mathcal{H}} ( H \rho ) = {\Tr}_{\mathcal{H}
( K )} ( H ( K ) \rho ( K ) ) + {\Tr}_{\mathcal{H} ( K ' )} ( H (
K ' ) \rho ( K ' )) = \ed \be = \overline{h} ( K ) + \overline{h}
( K' ) \, . \label{5.9} \ee
\end{quote}

\noindent Definition: Reduced State Operator of a Subsystem

\begin{quote}
Let $ \rho $ be the state operator of a composite system. The
operator \be \rho ( K ) = {\Tr}_{\mathcal{H} ( K ' )} ( \rho )
\label{5.10} \ee defined on the corresponding Hilbert space $
\mathcal{H} ( K ) $ is called the \emph{reduced state operator} of
subsystem $ K $.
\end{quote}

\noindent Definition: Independent State of a Subsystem

\begin{quote}
A subsystem $ K $ of a composite system is said to be in an
\emph{independent state} if, and only if, the state operator $
\rho $ of the composite system can be written as \be \rho = \rho (
K ) \otimes \rho ( K ' ) \label{5.11} \ee where $ \rho ( K ) $ is
the reduced state operator of subsystem $ K $ and $ \rho ( K ' ) $
is the reduced density operator of the ``complementary" subsystem
$ K ' $. Only in this case, the following relation holds \be \ln
\rho = \ln \rho ( K ) \otimes I ( K ' ) + I ( K ) \otimes \ln \rho
( K ' ) \, .  \label{5.12} \ee
\end{quote}

\noindent Definition: Independent Systems

\begin{quote}
A system is said to be \emph{independent} if, and only if, it is a
subsystem in an independent state of every conceivable composite
system which contains it.
\end{quote}

\noindent Corollary: Additivity of Entropy for Subsystems in
Independent States

\begin{quote}
The mean value of the entropy for a system composed of two
subsystems $ K $ and $ K ' $ in \emph{independent states} is equal
to the sum of the ``private entropy" values for the two
subsystems, i.e. \bd \overline{s} = - k {\Tr}_{\mathcal{H}} ( \rho
\ln \rho ) = - k {\Tr}_{\mathcal{H} ( K )} ( \rho ( K ) \ln \rho (
K ) ) - k {\Tr}_{\mathcal{H} ( K ' )} ( \rho ( K ' ) \ln \rho ( K
' ) ) = \ed \be = \overline{s} ( K ) + \overline{s} ( K ' ) \, .
\label{5.13} \ee In general (cf. E.H. Lieb and M.B. Ruskai
(1973)), the following inequality holds \be \overline{s} \leq
\overline{s} ( K ) + \overline{s} ( K ' ) \label{5.14} \ee where
the equal sign holds if, and only if, subsystem $ K $ is in an
independent state (this property of the entropy is called
\emph{subadditivity}).
\end{quote}

\noindent Definition: Isolated Systems

\begin{quote}
A system is said to be isolated if, and only if, it is separable
and independent.
\end{quote}

\noindent Since, as we will see, a separable system which is
initially independent remains independent indefinitely, this
definition implies that an isolated system must be independent at
the moment of its ``creation".

\section{NEW EQUATION OF MOTION FOR GENERAL QUANTUM DYNAMICS} \label{sez:5.2}

In view of the foregoing discussion and definitions we can now
present the equation of motion that we propose for Quantum
Thermodynamics. The equation of motion is introduced first for the
particular case of a system composed of a single elementary
constituent (eq. \ref{5.16}) and then for the case of a general
composite system (eq. \ref{5.30}).

\subsection{Equation of Motion for a Single Elementary Constituent} \label{sotsez:5.2.1}

Let the system of interest consist of only one elementary
constituent. The corresponding Hilbert space is $ \mathcal{H} $.
We have postulated that the corresponding list of generators of
the motion is

\be H , \, X , \, \ldots , \, Y \label{5.15} \ee

\noindent where we recall that the dimensionless generators $ X ,
\, ... , \, Y $, commute individually with the Hamiltonian
operator $ H $. We further postulate that to every type of
elementary constituent there corresponds a positive time constant
$ \tau $ entering explicitly its dynamical law of causal
evolution.

The proposed equation of motion for a single elementary
constituent is

\be \fr{d \rho}{d t} = - \, \fr{i}{\hbar} [H ,  \rho] -
\fr{\tau}{\hbar^2} \{D ,  \rho \} \label{5.16} \ee

\noindent where $ \{ \ , \ \} $ is the anticommutator symbol $ (
{F ,  G} = FG + GF ) $, $ D $ is an operator on $ \mathcal{H} $
defined by the determinant

\be D = \lat \begin{array}{ccccc} \Delta ( \ln \rho ) & \Delta H &
\Delta X & \ldots & \Delta Y \\ ( H ,  \ln \rho ) & ( H ,  H ) & (
H ,  X ) & \ldots & ( H ,  Y )  \\ ( X ,  \ln \rho ) & ( X ,  H )
& ( X ,  X ) & \ldots & ( X ,  Y )  \\ \ldots & \ldots & \ldots &
\ldots & \ldots \\ ( Y ,  \ln \rho ) & ( Y ,  H ) & ( Y ,  X ) &
\ldots & ( Y ,  Y ) \end{array} \rat \, , \label{5.17} \ee

\noindent and the symbols $ \Delta F $ and $ ( F ,  G ) $, for $ F
$ and $ G $ any Hermitian operator, are defined by

\be \Delta F = F - I \Tr ( \rho F ) \label{5.18} \ee

\noindent and

\be ( F ,  G ) = \Tr ( \rho \{ \Delta F ,  \Delta G \} ) \, .
\label{5.19} \ee

\noindent Note that $ ( F ,  G ) = ( G ,  F ) $ is a scalar
product on the real space $ \mathcal{L} $ of linear, Hermitian
operators on $ \mathcal{H} $. As we will see in the next Section,
the proposed equation of motion \ref{5.16} entails the
Schr\"odinger equation as a particular case, implies the existence
of irreversible processes for a single elementary constituent,
yields an explicit expression for the rate of entropy production
and entails a generalized statement of the second law of
Thermodynamics.

The experimental verification of the proposed equation of motion
for a single elementary constituent is clearly an open problem of
fundamental theoretical and practical importance. Moreover, the
identification of the non-Hamiltonian generators of the motion and
the time constant that are associated with every elementary
constituent, is still an open problem. However, the major theorems
that follow are independent of the specific form and number of
generators of the motion, and apply therefore to any elementary
constituent of matter.

\subsection{Theorems} \label{sotsez:5.2.2}

The major theorems of the proposed dynamics of a single elementary
constituent are as follows. Proofs are given in Appendix
\ref{app:C}.

\vspace{\baselineskip}

\noindent Th1QT

\begin{quote}
If the state operator is a projection operator $ \rho = P_{\psi}
$, onto a one-dimensional subspace of $ \mathcal{H} $, then the
equation of motion \ref{5.16} reduces to the Schr\"odinger
equation and the evolution of the isolated single elementary
constituent is therefore purely Hamiltonian.
\end{quote}

\noindent Definition: Constants of the Motion

\begin{quote}
A physical observable $ c $, represented by the functional $ c (
\rho ) $ defined on the quantal phase-domain of the system, is
said to be a \emph{constant of the motion} if, and only if, the
value of the representative functional is time invariant for any
state of the system.

If the observable $ c $ is represented by a linear functional,
then there exists a unique operator $ C $ such that, for all state
operators $ \rho$, $c ( \rho ) = \Tr  ( C \rho ) $. In this case
the observable is said to be a \emph{linear constant of the
motion} if, and only if, \be d \overline{c} ( \rho ) / dt = \Tr (
C ( d \rho / dt ) ) = 0 \label{5.20} \ee for all state operators.
\end{quote}

\noindent Th2QT

\begin{quote}
A physical observable represented by a linear functional $ c $ or,
equivalently, by a linear Hermitian operator $ C $, is a constant
of the motion if, and only if, $ C $ is a linear combination of
the generators of the motion and the identity operator, i.e. of $
I , \, H , \, X , \, ... , \, Y $.
\end{quote}

\noindent According to purely Hamiltonian dynamics (eq.
\ref{5.2}), any Hermitian operator which commutes with $ H $
represents a linear constant of the motion. Theorem Th2QT states
that not every linear constant of the motion according to
Hamiltonian dynamics is a constant of the motion according to
Quantum Thermodynamics. This is an important novel feature of the
proposed equation of motion.

\vspace{\baselineskip}

\noindent Definition: Equilibrium States

\begin{quote}
A state operator is said to represent an \emph{equilibrium state}
if, and only if, $ d \rho / d t = 0 $.
\end{quote}

\noindent Th3QT

\begin{quote}
The following conditions are necessary and sufficient   for
equilibrium:
\begin{description}
    \item[(a) ] the state operator $ \rho $ commutes with the
    Hamiltonian operator $ H $; and
    \item[(b) ] the state operator $ \rho $ commutes with a linear
    combination of the generators of the motion \begin{subequations}\be R = - \beta H
    + \chi X + \ldots + \upsilon Y \label{5.21a} \ee and the $ i
    $-th eigenvalue $ p_i $ of $ \rho $ is either \be p_i = 0
    \label{5.21b} \ee or \be p_i = ( 1 / z ) \exp ( R_i )
    \label{5.21c} \ee where \be z = \sum_i \exp ( R_i )
    \label{5.21d} \ee\end{subequations} and the index $ i $ runs over an eigenbasis
    shared by $ \rho $ and $ R $, whose existence is ensured by
    the fact that the two operators commute. Notice that state
    operators that are projection operators $ P_{\psi} $ onto
    one-dimensional subspaces of $ \mathcal{H} $ satisfying
    condition (a), also satisfy condition (b), the eigenvalues
    being all zero except for a single one equal to unity.
\end{description}
\end{quote}

\noindent Th4QT

\begin{quote}
The mean value of the entropy increases with time at a rate given
by the relation \be d \overline{s} / d t = k \tau g / \hbar^2 \geq
0 \label{5.22} \ee where $ g $ is a positive semi-definite
nonlinear functional defined by the following determinant \be g =
\lat \begin{array}{ccccc} ( \ln \rho ,  \ln \rho ) & ( \ln \rho ,
 H ) & ( \ln \rho ,  X ) & \ldots & ( \ln \rho ,  Y ) \\ ( H ,
\ln \rho ) & ( H ,  H ) & ( H ,  X ) & \ldots & ( H ,  Y )
\\ ( X ,  \ln \rho ) & ( X ,  H ) & ( X ,  X ) & \ldots & (
X ,  Y )  \\ \ldots & \ldots & \ldots & \ldots & \ldots \\ ( Y ,
\ln \rho ) & ( Y ,  H ) & ( Y ,  X ) & \ldots & ( Y ,  Y )
\end{array} \rat \, . \label{5.23} \ee The rate of entropy
production $ d \overline{s} / dt $ for a single elementary
constituent is zero if, and only if, the state is an equilibrium
state.
\end{quote}

\noindent This theorem is a statement of existence of irreversible
processes of any isolated physical system composed of a single
elementary constituent. Any system in any non-equilibrium state
will tend towards an equilibrium state with the same mean values
of the constants of the motion and a higher mean value of the
entropy.

\vspace{\baselineskip}

\noindent Definition: Stable Equilibrium States

\begin{quote}
An equilibrium state operator $ \rho $ is said to represent a
\emph{stable equilibrium state} if, and only if, there exist
\emph{no} other state operators with the same mean values of the
constants of the motion and a higher mean value of the entropy.
\end{quote}

\noindent Though we state it as a definition, we conjecture that
this statement should follow as a theorem of the proposed dynamics
and the mathematical definition of stability according to A.M.
Liapunov (1892) (cf. Appendix \ref{app:C}). The definition just
given can be justified as follows. Consider an equilibrium state
(as specified by Theorem Th3QT) and the ``plane" in quantal
phase-domain of state operators with the same values of the
constants of the motion. If on this ``plane" there exists a state
operator with a higher entropy than the equilibrium state under
consideration, then it is possible to find ``neighbouring" states
of the equilibrium state that are non-equilibrium and proceed
according to Theorem Th4QT towards states of higher entropy. Such
``neighbouring" states would leave the ``neighbourhood" of the
equilibrium state without ever being able to return to it, since
the entropy cannot decrease. Such an equilibrium state can
therefore be said to be unstable. An equilibrium state will be
stable only if it has the maximum entropy among the states with
the same mean values of the constants of the motion.

\vspace{\baselineskip}

\noindent Th5QT

\begin{quote}
Among the states of a system with given mean values of the
constants of the motion, one and only one is a stable equilibrium
state.
\end{quote}

\noindent This fundamental theorem is in fact a generalization of
a known statement of the second law of thermodynamics (cf. J.H.
Keenan, G.N. Hatsopoulos and E.P. Gyftopoulos (1972)).
Consequences of this theorem are discussed in Section
\ref{sotsez:5.2.4}.

\vspace{\baselineskip}

\noindent Th6QT

\begin{quote}
The explicit expression of the state operator for any elementary
constituent in a stable equilibrium state is
\begin{subequations}\be \rho = ( 1 / Z ) \exp ( - \beta H + \chi X
+ \ldots + \upsilon Y ) \label{5.24a} \ee where \be Z = \Tr ( \exp
( - \beta H + \chi X + \ldots + \upsilon Y ) \label{5.24b}
\ee\end{subequations} and the coefficients $ \beta , \, \chi , \,
... , \, \upsilon $ are determined by the initial values of a
complete and independent set of linear constants of the motion.
The function $ Z $ can be called the generalized \emph{partition
function}.
\end{quote}

\noindent Th7QT

\begin{quote}
The mean value of the entropy for an elementary constituent in a
stable equilibrium state can be written in terms of the mean
values of the generators of the motion and the partition function:
\be \overline{s} = k \beta \overline{h} - k \chi \overline{x} -
\ldots - k \upsilon \overline{y} + k \ln Z \, . \label{5.25} \ee
This relation can be called the generalized Gibbs equation for a
single elementary constituent.
\end{quote}

\subsection{Equation of Motion for a General Composite System} \label{sotsez:5.2.3}

Let us consider a general system composed of $ M $ distinguishable
elementary   constituents.   The corresponding   Hilbert space is
``structured" as follows

\bd \qquad \qquad \qquad \mathcal{H} = \bigotimes^{M}_{J = 1}
 \mathcal{H} ( J ) \, . \qquad \qquad \qquad
(\ref{5.6}) \ed

\noindent We postulate that the list of generators of the motion
for the $ J $-th elementary constituent is

\be V ( J ) , \, X ( J ) , \, \ldots , \, Y ( J ) \label{5.26} \ee

\noindent where the dimensionless operators $ X ( J ) , \, ... ,
\, Y ( J ) $ are the non-Hamiltonian generators of the motion
corresponding to a system composed of the $ J $-th elementary
constituent only. The operators $ V ( J ) $ are defined instead as
follows

\be V ( J ) = {\Tr}_{ \mathcal{H} ( J ' ) } ( ( I ( J ) \otimes
\rho ( J ' ) ) H ) \label{5.27} \ee

\noindent where $ \rho ( J ' ) $ is the reduced state operator of
the complementary subsystem of constituent $ J $, $ I  ( J ) $ is
the identity operator on the space $ \mathcal{H} ( J ) $ and $ H $
is the Hamiltonian operator of the overall composite system. The
operator $ V ( J ) $ may be called the \emph{reduced Hamiltonian
operator} of the $ J $-th elementary constituent, since for a
\emph{separable} constituent it reduces to

\be V ( J ) = H ( J ) + \overline{h} ( J ' ) \label{5.28} \ee

\noindent where

\be \overline{h} ( J ' ) = {\Tr}_{\mathcal{H} ( J ' )} ( H ( J ' )
\rho ( J ' ) ) \label{5.29} \ee

\noindent is the mean private energy of the complementary
subsystem $ J ' $ (cf. Section \ref{sotsez:5.1.5}, eq. \ref{5.9}).

The proposed equation of motion for a general composite system is

\be \fr{d \rho}{d t} = \hat{N} ( \rho ) = - \, \fr{i}{\hbar} [ H ,
 \rho ] - \fr{1}{\hbar^2} \sum_{J = 1}^M \tau ( J ) \{ D ( J ) ,
 \rho ( J ) \} \otimes \rho ( J ' ) \label{5.30} \ee

\noindent where $ \tau (J) $ is the positive time constant
characteristic of the $ J $-th elementary constituent, $ \rho ( J
) $ and $ \rho ( J ' ) $ are the reduced state operators of the $
J $-th elementary constituent and its complementary subsystem, the
operator $ D ( J ) $ on $ \mathcal{H} ( J ) $ is defined by the
determinant

\be D ( J ) = \lat \begin{array}{ccccc} \Delta W ( J ) & \Delta V
( J ) & \Delta X ( J ) & \ldots & \Delta Y ( J ) \\ ( V ( J ) , W
( J ) ) & ( V ( J ) ,  V ( J ) ) & ( V ( J ) ,  X ( J ) ) & \ldots
& ( V ( J ) ,  Y ( J ) )  \\ ( X ( J ) ,  W ( J ) ) & ( X ( J ) ,
 V ( J ) ) & ( X ( J ) ,  X ( J ) ) & \ldots & ( X ( J ) ,  Y ( J
) )  \\ \ldots & \ldots & \ldots & \ldots & \ldots
\\ ( Y ( J ) ,  W ( J ) ) & ( Y ( J ) ,  V ( J ) ) & ( Y ( J )
,  X ( J ) ) & \ldots & ( Y ( J ) ,  Y ( J ) ) \end{array} \rat \,
, \label{5.31} \ee

\noindent the symbols   $ \Delta F ( J ) $ and $ ( F ( J ) ,  G (
J ) ) $ ,   for $   F ( J ) $ and $ G ( J ) $ any Hermitian
operator on the factor Hilbert space $ \mathcal{H} ( J ) $, are
defined by

\be \Delta F ( J ) = F ( J ) - I ( J ) {\Tr}_{\mathcal{H} ( J )} (
\rho ( J ) F ( J ) ) \label{5.32} \ee

\noindent and

\be ( F ( J ) ,  G ( J ) ) = {\Tr}_{\mathcal{H} ( J )} ( \rho ( J
) \{ \Delta F ( J ) ,  \Delta G ( J ) \} ) \, , \label{5.33} \ee

\noindent and

\be W ( J ) = {\Tr}_{\mathcal{H} ( J ' )} ( ( I ( J ) \otimes \rho
( J ' ) ) \ln \rho ) \, . \label{5.34} \ee

\noindent It is noteworthy that, if the $ J $-th elementary
constituent is in an independent state, then the operator $ W ( J
) $ reduces to

\be W ( J ) = \ln \rho ( J ) - \overline{s} ( J ' ) / k
\label{5.35} \ee

\noindent where

\be \overline{s} ( J ' ) = - k {\Tr}_{\mathcal{H} ( J ' )} ( \rho
( J ' ) \ln \rho ( J ' ) ) \label{5.36} \ee

\noindent is the mean entropy of the complementary subsystem $ J '
$ (cf. Section \ref{sotsez:5.1.5}, eq. \ref{5.13}).

Equation \ref{5.30} defines explicitly the evolution superoperator
$ \hat{N} $ appearing in the postulates P5QT and P5QST, and
completes therefore the paradigms of Quantum Thermodynamics and
Quantum Statistical Thermodynamics.

Equation \ref{5.30} generalizes equation \ref{5.16} to systems
composed of any number $ M $ of distinguishable elementary
constituents. If equation \ref{5.16} can be written for each
elementary constituent, i.e. if the non-Hamiltonian generators of
the motion and the time constant associated with each elementary
constituent are given, then equation \ref{5.30} requires no
additional input. Thus, the solution of the open problems outlined
at the end of Section \ref{sotsez:5.2.1} implies the solution of
the general dynamical problem. Again, the major theorems that
follow are independent of the specific types of elementary
constituents and apply therefore to the general composite system.

\subsection{Theorems} \label{sotsez:5.2.4}

The major theorems of the proposed dynamics of a general composite
system are as follows. Proofs are given in Appendix \ref{app:C}.

\vspace{\baselineskip}

\noindent Th8QT

\begin{quote}
If the state operator of the general composite system is \be \rho
= \bigotimes^{M}_{J = 1}  P_{\psi} ( J ) \, , \label{5.37} \ee
where the index $ J $ runs over every elementary constituent,
then, and only then, the equation of motion \ref{5.30} reduces to
the Schr\"odinger equation of motion, and the evolution of the
isolated composite system is therefore purely Hamiltonian.
\end{quote}

\noindent It is noteworthy that, if the elementary constituents
are not all separable, then the initial state \ref{5.37} will
evolve into states of different type, eventually triggering the
non-Hamiltonian term in the equation of motion.

\vspace{\baselineskip}

\break

\noindent Th9QT

\begin{quote}
A physical observable represented by a linear functional $ c $ or,
equivalently, by a linear Hermitian operator $ C $, is a constant
of the motion for the general composite system if, and only if,
\begin{description}
    \item[(a) ] $ [ C ,  H ] = 0 $ and
    \item[(b) ] $ C $ is a linear combination of the   operators
    \bd I , \, H , \, X ( 1 ) \otimes I ( 1 ' ) , \, \ldots , \, Y
    ( 1 ) \otimes I ( 1 ' ) , \, X ( 2 ) \otimes I ( 2 ' ) , \,
    \ldots , \, Y ( M ) \otimes I ( M ' ) \ed where $ I ( J ) , \,
    X ( J ) , \, ... , \, Y ( J ) $ are the identity operator and
    the non-Hamiltonian generators of the motion of the $ J $-th
    elementary constituent and $ I , \, H $ are the identity
    operator and the Hamiltonian operator of the overall composite
    system. It follows that the physical observable represented by
    the overall Hamiltonian operator $ H $ is always a constant of
    the motion for any isolated physical system. This physical
    observable is called the \emph{energy} of the system.
\end{description}
\end{quote}

\noindent Th10QT

\begin{quote}
The mean value of the entropy increases with time  at a rate given
by the relation \be d \overline{s} / d t = \sum_J k\, \tau ( J ) g
( J ) / \hbar^2 \geq 0 \label{5.38} \ee where $ g ( J ) $ is a
positive semi-definite nonlinear functional defined by the
following determinant \be g ( J ) = \lat \begin{array}{ccccc} ( W
( J ) ,  W ( J ) ) & ( W ( J ) ,  V ( J ) ) & ( W ( J ) ,  X ( J )
) & \ldots & ( W ( J ) ,  Y ( J ) ) \\ ( V ( J ) ,  W ( J ) ) & (
V ( J ) ,  V ( J ) ) & ( V ( J ) ,  X ( J ) ) & \ldots & ( V ( J )
, Y ( J ) )  \\ ( X ( J ) ,  W ( J ) ) & ( X ( J ) ,  V ( J ) ) &
( X ( J ) ,  X ( J ) ) & \ldots & ( X ( J ) ,  Y ( J ) )
\\ \ldots & \ldots & \ldots & \ldots & \ldots
\\ ( Y ( J ) ,  W ( J ) ) & ( Y ( J ) ,  V ( J ) ) & ( Y ( J )
, X ( J ) ) & \ldots & ( Y ( J ) ,  Y ( J ) ) \end{array} \rat \,
. \label{5.39} \ee
\end{quote}

\noindent This theorem is a statement of existence  of
irreversible processes for a general isolated physical system. Any
such system in any non-equilibrium state tends towards an
equilibrium state with the same mean values of the constants of
the motion and a higher mean value of the entropy.

\vspace{\baselineskip}

\noindent Th5QT

\begin{quote}
Among the states of a system with given mean values  of the
constants of the motion, one and only one is a stable equilibrium
state.
\end{quote}

\noindent As noted in Section \ref{sotsez:5.2.2},  this is a
generalized statement of the second law of Thermodynamics. When
the constants of the motion are the energy and the number of
particles of constituent species, this theorem coincides with the
fourth postulate of the unified quantum theory of Mechanics and
Thermodynamics proposed by Hatsopoulos and Gyftopoulos (1976). The
first three postulates of their unified quantum theory are
equivalent to postulates P1QT through P4QT of Quantum
Thermodynamics. Thus, the theorems proved within that theory from
the four postulates and the underlying definitions, are also
theorems of Quantum Thermodynamics, if the same definitions are
adopted.

\vspace{\baselineskip}

\break

\noindent Th11QT

\begin{quote}
The explicit expression of the state operator for a general
composite system in a stable equilibrium state is \be \rho = ( 1 /
Z ) \exp ( - \beta H + \gamma_1 C_1 + \ldots + \gamma_N C_N )
\label{5.40} \ee where \be Z = \Tr ( \exp ( - \beta H + \gamma_1
C_1 + \ldots + \gamma_N C_N ) ) \label{5.41} \ee and the operators
$ H , \, C_1 , \, ... , \, C_N $  form a complete and independent
set of linear constants of the motion for the composite system.
The coefficients $ \beta , \, \gamma_1 , \, ... , \, \gamma_N $
are determined by the initial values of the constants of the
motion. The function $ Z $ can be called the generalized partition
function.
\end{quote}

\noindent Th12QT

\begin{quote}
The mean value of the entropy of a general composite system in any
stable equilibrium state can be written in terms of the mean
values of a complete and independent set of constants of the
motion and the partition function \be \overline{s} = k \beta
\overline{h} - k \gamma_1 \overline{c}_1 - \ldots - k \gamma_N
\overline{c}_N + k \ln Z \, . \label{5.42} \ee This relation can
be called the  generalized Gibbs equation.
\end{quote}

\noindent The following theorems  refer to subsystems of the
general composite system. The notation introduced in Section
\ref{sotsez:5.1.5} is implied.

\vspace{\baselineskip}

\noindent Th13QT

\begin{quote}
The causal evolution of the reduced  state operator of subsystem $
K $ (with elementary constituents indexed by $ J ( K ) $) is
determined via the equation of motion (reduced equation of motion)
\be \fr{d \rho ( K )}{d t} = - \, \fr{i}{\hbar} {\Tr}_{\mathcal{H}
( K ' )} ( [ H ,  \rho ] ) - \, \fr{1}{\hbar^2} \sum_{J ( k ) =
1}^{M ( K )} \tau ( J ( K ) ) \{ D ( J ( K ) ) ,  \rho ( J ( K ) )
\} \otimes \rho ( J ' ( K ) ) \, . \label{5.43} \ee
\end{quote}

\noindent Th14QT

\begin{quote}
Given the reduced equations of motion for the two subsystems $ K $
and $ K ' $, the equation of motion of the composite system is
given by \be \fr{d \rho}{d t} = \fr{d \rho ( K )}{d t} \otimes
\rho ( K ' ) + \rho ( K ) \otimes \fr{d \rho ( K ' )}{d t}
\label{5.44} \ee if, and only if, the two subsystems are
\emph{separable} and in \emph{independent} states.
\end{quote}

\break

\noindent Th15QT

\begin{quote}
The rate of entropy production of a system composed of two
subsystems $ K $ and $ K ' $ in \emph{independent} states equals
the sum of the corresponding rates of private entropy production
if, and only if, the two subsystems are \emph{separable}. Then,
and only then, \be d \overline{s} / dt = d \overline{s} ( K ) / d
t + d \overline{s} ( K ' ) / d t \, . \label{5.45} \ee Thus, two
\emph{separable} systems that are initially  \emph{independent}
remain independent indefinitely.
\end{quote}

\noindent We conclude with some qualitative remarks on the  notion
of reversibility. By virtue of theorem Th10QT, the equation of
motion \ref{5.30} is fundamentally irreversible. If the time
constants $ \tau $ are nonzero for ``all" elementary constituents,
then no material system evolves according to pure Hamiltonian
dynamics and there exists no material system with an inherent
dynamical behaviour which is purely reversible. Reversible
processes, however, can be conceived as the result of either
special initial conditions or special limiting situations. For
example, eq. \ref{5.37} identifies a special initial condition for
which the evolution is initially purely Hamiltonian. Special
limiting situations in which the reversible behaviour is
approximated, may be verified when the interaction between the
constituents of the system are so ``strong" that the Hamiltonian
term in the equation of motion dominates and overwhelms, for a
``short" time, the effect of the irreversible non-Hamiltonian
term. Again, for states that are very ``close" to stable
equilibrium, the non-Hamiltonian contribution may be so small that
its ``linearization" may become a reasonable approximation. A
promising ground is immediately open for a fundamental
justification of the successful phenomenological theory of linear
irreversible phenomena based on the well known Onsager relations.

These qualitative comments show that the proposed equation of
motion has opened several new avenues for further fundamental
studies, the theoretical and practical importance of which need
not be mentioned.

\chapter{SUMMARY AND RECOMMENDATIONS} \label{cap:6}

The purpose of this dissertation has been to present a general
quantum theory in which two fundamental problems of contemporary
quantum physics find a consistent solution: the lack of a complete
equation of motion and the lack of an unambiguous quantum
statistical theory.

Quantum Thermodynamics, the physical part of the proposed theory,
resolves the incompleteness of the current dynamical theory by
introducing a new nonlinear equation of motion (eq. \ref{5.30}).
The major implications of Quantum Thermodynamics can be summarized
as follows: the proposed equation of motion reduces to the
Schr\"odinger equation of motion for a particular class of states;
among the states of a system with given mean values of the
constants of the motion there exist many equilibrium states,
however, one and only one is a stable equilibrium state (this is a
generalization of a known statement of the second law of
Thermodynamics); any isolated system in any non-equilibrium state
evolves towards states of higher entropy (this is a statement of
existence of irreversible processes for isolated systems). It is
worth stressing that Quantum Thermodynamics is concerned only with
a description of the states of any individual physical system and
their causal evolution. It is not a statistical theory. In Quantum
Thermodynamics, the laws of Thermodynamics and irreversibility
emerge as exact consequences of the fundamental quantum
\emph{dynamical} behaviour of the elementary constituents of any
individual material system, microscopic or macroscopic, simple or
complex. This general statement is a major conclusion of the
dissertation.

Quantum Statistical Thermodynamics, the statistical part of the
proposed theory, resolves the present lack of an unambiguous
quantum statistical theory by adopting a new measure-theoretic
description of ensembles. This statistical theory is superimposed
to and consistent with the laws of Quantum Thermodynamics. It is
concerned with the description of stochastic distributions of
states in an ensemble of identical systems, each of which
individually obeys the laws of Quantum Thermodynamics. To every
such ensemble, Quantum Statistical Thermodynamics associates a
measure-theoretic descriptor, called the statistical-weight
measure, which represents the distribution of individual quantum
thermodynamical states in the ensemble. The proposed description
is unambiguous in that, for any given ensemble, it uniquely
identifies both the spectrum of states represented in the ensemble
and their relative population. In other words, it accounts
unambiguously for the essential distinction between quantal and
nonquantal uncertainties.

\vspace{\baselineskip}

Many interesting problems arise from the present work and require
further investigation.

\vspace{\baselineskip}

The new equation of motion proposed for Quantum Thermodynamics
opens a large number of questions and routes of theoretical and
experimental investigation. First of all, the non-Hamiltonian
generators of the motion and the time constant associated with
each type of elementary constituent have been left unspecified.
Their specification is an open problem the solution of which will
probably require a constructive confrontation between conventional
quantum theory, the present theory and the results of past and
newly designed experiments. This confrontation will also decide
upon the validity of the proposed equation to describe
irreversible phenomena. Other theoretical problems consist of
deriving from Quantum Thermodynamics more detailed theorems that
will help to clarify the physical significance of the underlying
nonlinear dynamics. For example, the description of known
reversible processes of the type mentioned in Section
\ref{sotsez:5.1.1} point (b), is still an unsolved problem. We
have only conjectured that reversible processes may be suitably
approximated under special limiting conditions, but we have not
explicitly identified such conditions.

The methods of the so-called linear irreversible phenomenological
thermodynamics, based on the well known Onsager reciprocity
relations, are known to successfully describe a certain class of
irreversible phenomena. Also the Boltzmann equation of the kinetic
theory of gases is known to describe successfully a certain class
of irreversible processes. Yet, both the Onsager relations and the
Boltzmann equation cannot be rigorously derived from conventional
Hamiltonian dynamics, since that dynamics is fundamentally
reversible. The proposed nonlinear irreversible dynamics is
instead fundamentally irreversible, and holds the promise to
provide a unique fundamental ground for a rigorous derivation of
the Onsager relations and the Boltzmann equation.

Finally, application of the proposed dynamical theory to practical
problems of physics and engineering is at this moment a most
challenging prospect.

As regards Quantum Statistical Thermodynamics, the novel
measure-theoretic description of ensembles or, equivalently, of
preparations, has introduced two problems. The first problem
concerns the empirical determination of the statistical-weight
measure describing the preparation under consideration. We have
conjectured a possible form of a ``quorum" of physical observables
the experimental determination of which might prove sufficient to
determine uniquely the statistical-weight measure (Section
\ref{sotsez:4.2.3}). However, even if this conjecture or a
generalization of it were proved, the important physical problem
to attach a definite physical and operational meaning to the
observables in such a ``quorum" would still remain a challenge.
The second problem pertains to the domain of Information Theory.
We have studied, in the context of Quantum Statistical
Thermodynamics, three typical problems of Information Theory and
found satisfactory answers only for the particular case in which
the spectrum of the statistical-weight measure is known and
countable. The more general case in which the spectrum is unknown
or noncountable leads to difficulties and requires further
analysis.

\appendix

\chapter{AMBIGUOUS DESCRIPTIONS OF HETEROGENEOUS PREPARATIONS} \label{app:A}

This Appendix reports two unsuccessful attempts to obtain a
mathematical description of heterogeneous preparations reflecting
unambiguously the concept of state of an individual system. The
first, based on a description by means of a statistical operator,
is found not to satisfy both condition (ii) and (iii) of Section
\ref{sotsez:3.2.3}. The second, based on the description by means
of a statistical superoperator, is found not to satisfy condition
(iii).

\section{Description by means of a Statistical Operator} \label{sez:A.1}

Consider a heterogeneous preparation, defined by the symbolic
expression

\bd \Pi = \sum_n w_n \Pi_n \ed

\noindent where the component preparations $ \Pi_n $ are
homogeneous. Let us attempt to describe such a preparation by
means of a statistical operator $ W $ (linear, Hermitian,
nonnegative-definite and unit-trace) defined on the Hilbert space
of the system. To do so we can adopt postulates P2QSM and P3QSM of
Section \ref{sotsez:3.2.2}, so that the statistical operator $ W $
is linked to the expected mean values of physical observables
according to the relation

\bd \qquad \qquad \qquad \langle  \overline{g} \rangle  = \Tr ( W
G ) \, . \qquad \qquad \qquad (\ref{3.11}) \ed

\noindent Because the structure of the given heterogeneous
preparation is specified, the \emph{expected mean value} $ \langle
\overline{g} \rangle  $ must equal the weighted sum of \emph{mean
values} $ { \overline{g} }_n $ corresponding to the homogeneous
component preparations $ \Pi_n $. This condition, i.e.

\be \Tr ( W G ) = \langle  \overline{g} \rangle  = \sum_n w_n {
\overline{g}_n } = \sum_n w_n \Tr ( \rho_n G ) \, , \label{A.1}
\ee

\noindent is satisfied for all operators $ G $ if, and only if,

\be W = \sum_n w_n \rho_n \label{A.2} \ee

\noindent Equation \ref{A.2} defines the rule of composition
required by condition (i) of the Section \ref{sotsez:3.2.3}.
However, the description of preparations by means of a statistical
operator leads to two types of ambiguities.

The first ambiguity arises from the failure to satisfy condition
(ii). The set of irreducible statistical operators (the idempotent
statistical operators) is in one-to-one correspondence to the set
of idempotent state operators. Thus, the set of irreducible
statistical operators is not mathematically rich enough to
identify unambiguously all the distinct homogeneous preparations
contemplated in Quantum Thermodynamics, i.e. all state operators.
In other words, a statistical theory superimposed to Quantum
Thermodynamics and based on such a description of preparations
could not reproduce the rule of correspondence between states and
homogeneous preparations.

The second ambiguity arises from the failure to satisfy condition
(iii). This is due to the mathematical fact that the statistical
operator $ W $ can be resolved into an infinity of weighted sums
of \emph{different} state operators, such as

\be W= \sum_n w_n \rho_n = \sum_q w_q' \rho_q' = \ldots
\label{A.3} \ee

\noindent The ambiguity concerns the notion of state of an
individual system. From the above multiple decomposition and
postulate P4QT of Quantum Thermodynamics (Section
\ref{sotsez:4.1.1}), a logical conclusion would be that an
individual system prepared according to the scheme $ \Pi $ is, for
example, in state $ \rho_n $ with probability $ w_n $ and, at the
same time, in state $ \rho_q' $ with probability $ w_q' $.
Analogously to what seen in Section \ref{sotsez:3.2.2}, such a
system would be a \emph{quantum monster}, i. e. a \emph{single}
individual system which is concurrently "in" \emph{two different}
states.

We conclude that the mathematical description of heterogeneous
preparations by means of a statistical operator is inadequate,
inconsistent and ultimately leads to paradoxes.

\section{Description by means of a Statistical Superoperator} \label{sez:A.2}

Next we consider an unsuccessful, but instructive, attempt to find
an unambiguous mathematical description of heterogeneous
preparations. The approach is based on an extension of the
axiomatic formulation of Quantum Statistical Mechanics to the next
higher mathematical level. Unfortunately, such formulation is
found to be ambiguous in that condition (iii) of Section
\ref{sotsez:3.2.3} is not satisfied. In fact, the approach is
based on the erroneous hypothesis that the paradigm of Quantum
Statistical Mechanics (that we reviewed in Section
\ref{sotsez:3.2.2}) is a correct way to superimpose a statistical
theory to Quantum Mechanics.

The approach consists of proceeding by mathematical analogy and
superimpose a statistical theory to Quantum Thermodynamics. Let $
\mathcal{H} $ be the Hilbert space of the system of interest. Let
$ \mathcal{L} $, be the \emph{real} space of linear, Hermitian
operators on $ \mathcal{H} $. Points (or "vectors") in this space
are operators on $ \mathcal{H} $. A state operator on $
\mathcal{H} $, the state descriptor in Quantum Thermodynamics, is
therefore a vector in $ \mathcal{L} $. This is analogous to the
fact that the state descriptor in Quantum Mechanics is a vector in
$ \mathcal{H} $. The analogy then leads to assume that linear,
simmetric operators $ \overline{G} , \, \overline{H} , \, \ldots $
defined on $ \mathcal{L} $ (superoperators on $ \mathcal{H} $)
correspond to physical observables of the system and, more
importantly, that the mathematical correspondent of a generic
preparation scheme $ \Pi $ is a linear, symmetric,
nonnegative-definite, unit-(super)trace operator $ \hat{W} $
defined on $ \mathcal{L} $: the direct analogous of the
statistical operator $ W $ of Quantum Statistical Mechanics. The
operator $ \hat{W} $ (a superoperator on $ \mathcal{L} $) would be
called the \emph{statistical superoperator} and would be linked to
the expected mean value $ \langle  \overline{g} \rangle $ of an
observable $ \hat{G} $ via evaluation of the (super)trace
functional

\be \langle  \overline{g} \rangle  = \hat{\Tr} ( \hat{W} \hat{G} )
\, . \label{A.4} \ee

An analysis of this mathematical description in the light of the
conditions of Section \ref{sotsez:3.2.3} is in order. Condition
(i) is satisfied by defining the following rule of composition of
different statistical superoperators

\be \hat{W} = \sum_n w_n { \hat{W} }_n \, . \label{A.5} \ee

\noindent Condition (ii) is also satisfied, and this is a partial
success of the approach. The set of statistical superoperators
contains in fact irreducible elements that are not decomposable
into a weighted sum of \emph{different} statistical
superoperators. The typical irreducible element is an idempotent
statistical superoperator $ \hat{P}_A $, i.e. a projection
(super)operator onto the one-dimensional subspace of $ \mathcal{L}
$ spanned by operator $ A $. The class of irreducible elements is
mathematically rich enough to describe all the possible
homogeneous preparations, since it is possible to define the
correspondence

\be \rho = A^2 \label{A.6} \ee

\noindent from the set of irreducible statistical superoperators
to the set of state operators.

The description fails however to satisfy condition (iii) and is
therefore ambiguous. The ambiguity originates from the fact that a
non-idempotent statistical superoperator $ \hat{W} $ can be
resolved into an infinity of weighted sums of \emph{different}
sets of projection superoperators, such as

\be \hat{W} = \sum_n w_n \hat{P}_{A_n} = \sum_q w_q' \hat{P}_{A_q}
= \ldots \label{A.7} \ee

\noindent The reasons why the non-uniqueness of resolution of the
mathematical descriptor of a heterogeneous preparation leads to
ambiguities have been already analysed in the last Section and in
Section \ref{sotsez:3.2.2}.

We conclude that the mathematical description of heterogeneous
preparations by means of a statistical superoperator $ \hat{W} $
is inconsistent and ultimately leads to paradoxes.

\chapter{GENERALIZED LINEAR DYNAMICS} \label{app:B}

This Appendix considers possible generalizations of Hamiltonian
dynamics based on equations of motion of the type

\bd \qquad \qquad \qquad \fr{d \rho}{d t} = \hat{L} ( \rho )
\qquad \qquad \qquad (\ref{5.5}) \ed

\noindent where the evolution superoperator $ \hat{L} $ is linear
in the state operator, i.e.

\be \hat{L} ( a_1 \rho_1 + a_2 \rho_2 ) = a_1 \hat{L} ( \rho_1 ) +
a_2 \hat{L} ( \rho_2 ) \label{B.1} \ee

\noindent for all state operators $ \rho_1 , \, \rho_2 $ and for
all scalars $ a_1 , \, a_2 $.

\section{Kossakowski-Lindblad Linear Superoperator} \label{sez:B.1}

In order to qualify as a bona fide evolution superoperator, $
\hat{L} $ must satisfy a number of necessary and sufficient
conditions that have first been stated by A. Kossakowski (1972a).
Kossakowski (1972b), for a special case, and G. Lindblad (1976),
for the general case, have shown that the most general linear
superoperator $ \hat{L} $ satisfying such conditions is

\be \hat{L} ( \rho ) = \hat{L}_o ( \rho ) + \sum_j \hat{L}_j (
\rho ) \label{B.2} \ee

\noindent where

\begin{subequations}
\be \hat{L}_o ( \rho ) = i [ B ,  \rho ] \, , \label{B.3a} \ee \be
\hat{L}_j ( \rho ) = A_j^{\dagger} \rho A_j - \, \fr{1}{2} \{
A_j^{\dagger} A_j ,  \rho \} \, , \label{B.3b}
\ee\end{subequations}

\noindent the operator $ B $ is Hermitian and the operators $ A_j
$ ($ A_j^{\dagger} $ indicates the Hermitian conjugate) are
instead not necessarily Hermitian. We call $ \hat{L} $ the $ K-L $
(Kossakowski-Lindblad) superoperator.

For the purposes of Quantum Thermodynamics, this form of the $ K-L
$ superoperator is still too unrestricted. Since the object of
study of Quantum Thermodynamics is the isolated system, we must at
least require that the mean value of the physical observable
energy be conserved and the mean value of the physical observable
entropy be never decreasing.

\section{Energy Conservation and Entropy Production} \label{sez:B.2}

The condition that the mean value of the energy of an isolated
system be conserved, imposes restrictions on the choice of the
operators $ B $ and $ A_j $ appearing in the definition of the $
K-L $ superoperator. The condition is that, for any state
operator,

\be d \overline{h} / d t = \Tr ( H ( d \rho / d t ) ) = \Tr ( H
\hat{L} ( \rho ) ) = 0 \label{B.4} \ee

\noindent where $ H $ is the Hamiltonian operator of the system.
By substituting the expression for $ \hat{L} $, we find that the
condition is satisfied if, and only if,

\be i [ H ,  B ] + \sum_j [ A_j ,  H A_j^{\dagger} ] = 0 \, .
\label{B.5} \ee

\noindent An obvious choice for $ B $ is $ B = - H / \hbar $, so
that the superoperator $ \hat{L}_o $ becomes the Liouvillian (eq.
\ref{5.3}). However, the choice of the operators $ A_j $ is not
obvious, expecially in view of the next condition.

The condition that the mean value of the entropy of an isolated
system be never decreasing imposes restrictions on the choice of
the operators $ A_j $ defining the $ K-L $ superoperator. The
condition is that, for any state operator,

\be d \overline{s} / dt = - k \Tr ( ( d \rho / dt ) \ln \rho ) = -
k \Tr ( \hat{L} ( \rho ) \ln \rho ) \geq 0 \, . \label{B.6} \ee

\noindent By substituting the expression for $ \hat{L} $, we find
that the condition becomes

\begin{subequations}
\be d \overline{s} / dt = k \sum_j \Tr ( A_j^{\dagger} A_j \rho
\ln \rho -  A_j^{\dagger} \rho A_j \ln \rho ) \geq 0 \label{B.7a}
\ee

\noindent or

\be d \overline{s} / dt =  k \sum_{j i n} { ( A_j^{\dagger} ) }_{i
n} { ( A_j ) }_{n i} ( \rho_i - \rho_n ) \ln \rho_i \geq 0
\label{B.7b} \ee\end{subequations}

\noindent where we have expanded the first expression with respect
to an eigenbasis of the state operator. Since the operators $ A_j
$ cannot depend on the state operator (by the linearity of the $
K-L $ superoperator), it follows that for any \emph{singular}
state operator (i.e., such that at least one eigenvalue is zero),
the rate of entropy production is \emph{infinite}. We find this
feature of any linear equation of this type undesirable, since it
would imply that a large class of states (including all the states
considered in Quantum Mechanics) are highly unstable and evolve
towards more stable states at an infinite rate. This observation,
added to the general observation made by Simmons about the
inadequacy of linear dynamics (cf. Section \ref{sotsez:5.1.2}),
has led us to consider nonlinear evolution superoperators.
However, an interesting example of linear equation of motion is
noteworthy.

\section{Pauli Master Equation} \label{sez:B.3}

A particular choice of the operators $ A_j $, which in general
does not satisfy the conditions just specified, leads to the
famous Pauli master equation. Let us substitute the index $ j $
with the double index $ rs $ and let

\be A_j = A_{r s} = c_{r s} | r \rangle  \langle  s | \qquad
\qquad ( A_{r s}^{\dagger} = c_{r s}^* | r \rangle  \langle  s | )
\label{B.8} \ee

\noindent where the vectors $ | i \rangle  $ form an eigenbasis
belonging to the Hamiltonian operator, $ | i \rangle  \langle  j |
$ are the corresponding dyadic operators and $ c_{rs} $ are
complex scalars. The corresponding equation of motion becomes

\be \fr{d \rho}{d t} = - \, \fr{i}{\hbar} [ H ,  \rho ] + \sum_{r
s} c_{rs} c_{rs}^* ( \rho_{r r} | s \rangle  \langle  s | - \,
\fr{1}{2} \{ | s \rangle  \langle  s |  , \rho \} ) \, .
\label{B.9} \ee

\noindent Letting $ c_{rs} c_{rs}^* = w_{r s} $ (clearly, $ w_{r
s} \geq 0 $) and taking the $ ij $-th matrix element of the
equation with respect to the same Hamiltonian eigenbasis, we
obtain the master equation

\be \fr{D \rho_{i j}}{D t} = \delta_{i j} \sum_r w_{i r} \rho_{r
r} - \rho_{i j} \fr{1}{2} \sum_r ( w_{r i} + w_{r j} )
\label{B.10} \ee

\noindent where we defined

\be \fr{D \rho_{i j}}{D t} = \fr{d \rho_{i j}}{d t} +
\fr{i}{\hbar} \rho_{i j} ( E_i - E_j ) \, . \label{B.11} \ee

\noindent For state operators that commute with the Hamiltonian $
H $, the eigenvalues $ p_i $ of the state operator vary according
to the equation

\be \fr{d p_i}{d t} = \sum_r w_{i r} p_r - p_i \sum_r w_{r i} \, .
\label{B.12} \ee

\noindent In order for the Pauli master equation to be energy
conserving, the following condition must be satisfied

\be \sum_j [ A_j ,  H A_j^{\dagger} ] = \sum_{r s} w_{r s} E_s ( |
r \rangle  \langle  r | - | s \rangle  \langle  s | ) = 0
\label{B.13} \ee

\noindent or, by taking matrix elements, the condition

\be \sum_s ( w_{n s} E_s - w_{s n} E_n ) = 0 \label{B.14} \ee

\noindent must be satisfied for every $ n $. The entropy
production condition is

\be d \overline{s} / d t = k \sum_{r s} w_{r s} ( \rho_{r r} -
\rho_{s s} ) \ln \rho_{s s} \geq 0 \, . \label{B.15} \ee

\noindent We see that the energy conservation and entropy
production requirements impose very restrictive conditions on the
coefficients $ w_{r s} $. If we set $ w_{r s} = w_{s r} $, to
satisfy the entropy condition, then the energy condition imposes
that $ w_{r s} = w \delta_{r s} $. The equation of motion then
becomes

\be \fr{D \rho_{i j}}{D t} = w ( \delta_{i j} \rho_{i i} - \rho_{i
j} ) \label{B.16} \ee

\noindent or

\be \fr{D \rho}{D t} = w \left( \sum_i | i \rangle  \rho_{i i}
\langle i | - \rho \right) \, . \label{B.17} \ee

\noindent This equation satisfies the energy conservation and
entropy production requirements, however, it is no exception to
Simmon's observation and the observation about infinite entropy
production for singular state operators.

To conclude, we present another example of linear equation which
satisfies the energy and the entropy conditions. Let the set of
operators $ A_j $ be composed of the Hermitian operator $ i
\tau^{\fr{1}{2}} F / \hbar $ only, where $ \tau $ is a positive
time constant and $ [ F ,  H ] = 0 $. The corresponding equation
is

\be \fr{d \rho}{d t} = - \, \fr{i}{\hbar} [ H ,  \rho ] -
\fr{\tau}{2 \hbar^2} [ F ,  [ F ,  \rho ] ] \, . \label{B.18} \ee

\noindent This equation has already been considered by Park and
Band (1978).

\chapter{PROOFS OF MAJOR THEOREMS OF QUANTUM THERMODYNAMICS} \label{app:C}

This Appendix outlines the proofs of the major theorems of Quantum
Thermodynamics that are enounced in Chapter \ref{cap:5}.

\vspace{\baselineskip}

\noindent Th1QT

\noindent Let the state operator be a projection operator and
substitute into eq. \ref{5.16}. The first term of the equation
results in eq. \ref{5.1} which is totally equivalent to the
Schr\"odinger equation of motion. The second term of eq.
\ref{5.16} vanishes. To see this, we observe that the operator $
P_{\psi} \ln P_{\psi} $ is the null operator, therefore, by the
definition \ref{5.19}, all the scalar products in the first column
of \ref{5.17} vanish. The first entry of the column does not
vanish but becomes the null operator when $ D $ is inserted into
eq. \ref{5.16}.

\vspace{\baselineskip}

\noindent Th2QT

\noindent The rate of change of the mean value of an observable
represented by the Hermitian operator $ C $, i.e.

\be \Tr ( C ( d \rho / dt ) ) \, , \label{C.1} \ee

\noindent is given by

\be ( - i / \hbar ) \Tr ( \rho  [ H ,  C ] ) - ( \tau / \hbar^2 )
\lat \begin{array}{ccccc} ( C ,  \ln \rho ) & ( C ,  H ) & ( C , X
) & \ldots & ( C ,  Y ) \\ ( H ,  \ln \rho ) & ( H ,  H ) & ( H ,
X ) & \ldots & ( H ,  Y )  \\ ( X ,  \ln \rho ) & ( X ,  H ) & ( X
,  X ) & \ldots & ( X ,  Y )  \\ \ldots & \ldots & \ldots & \ldots
& \ldots \\ ( Y ,  \ln \rho ) & ( Y ,  H ) & ( Y ,  X ) & \ldots &
( Y ,  Y ) \end{array} \rat \label{C.2} \ee

\noindent where we substituted eq. \ref{5.16} into eq. \ref{C.1}
and used eqs. \ref{5.18} and \ref{5.19}. For $ C $ to represent a
linear constant of the motion, eq. \ref{C.2} must vanish for any
state operator. The first term in the equation requires that $ C $
commute with $ H $. In order for the determinant in the second
term of eq. \ref{C.2} to vanish for any state operator it is
necessary and sufficient that a row (column) is either a linear
combination of the other rows (columns) or else it vanishes. By
the definitions \ref{5.18} and \ref{5.19}, the first row vanishes
if $ C $ is proportional to the identity operator $ I $. Since the
generators $ H , \, X , \, ... , \, Y $ are fixed, the only other
possibility is that the first row is a linear combination of the
other rows. This occurs independently of the state operator only
if $ C $ is a linear combination of the generators of the motion
(recall that $ ( \ , \ ) $ is a real scalar product and is
therefore linear in each factor). Thus, the second term of eq.
\ref{C.2} vanishes if and only if

\be C = a I + b H + c X + \ldots + d Y \, . \label{C.3} \ee

\noindent Since all the generators of the motion commute with $ H
$, also the first term vanishes. The possibility that the
difference between the first term and the second term in \ref{C.2}
vanishes independently of the state operator, while the two terms
do not vanish individually, is excluded. In fact, with respect to
the state operator, the first term is a linear functional while
the second is a nonlinear functional, therefore the two
functionals cannot be identical independently of the state
operator.

\vspace{\baselineskip} \noindent Th3QT

\noindent The first term of eq. \ref{5.16} vanishes if and only if
the state operator commutes with $ H $. Let us assume this is the
case. The second term becomes proportional to the operator

\be Q = \lat \begin{array}{ccccc} \{ \rho ,  \Delta \ln \rho \} &
\{ \rho ,  \Delta H \} & \{ \rho ,  \Delta X \} & \ldots & \{ \rho
, \Delta Y \} \\ ( H ,  \ln \rho ) & ( H ,  H ) & ( H ,  X ) &
\ldots & ( H ,  Y )  \\ ( X ,  \ln \rho ) & ( X ,  H ) & ( X , X )
& \ldots & ( X ,  Y )  \\ \ldots & \ldots & \ldots & \ldots &
\ldots \\ ( Y ,  \ln \rho ) & ( Y ,  H ) & ( Y ,  X ) & \ldots & (
Y ,  Y ) \end{array} \rat \, . \label{C.4} \ee

\noindent By the properties of determinants, $ Q $ is not altered
if we subtract from the first column a linear combination of the
other columns. We do so by using the same coefficients of eq.
\ref{5.21a}. By recalling the definition of $ (\  , \ ) $, we
obtain

\be Q = \lat \begin{array}{cccc} \{ \rho ,  \Delta \ln \rho -
\Delta R \} & \{ \rho ,  \Delta H \} & \ldots & \{ \rho , \Delta Y
\} \\ \Tr ( \Delta H \{ \rho ,  \Delta \ln \rho - \Delta R \} ) &
( H ,  H ) & \ldots & ( H ,  Y )  \\ \Tr ( \Delta X \{ \rho ,
\Delta \ln \rho - \Delta R \} ) & ( X ,  H ) & \ldots & ( X ,  Y )
\\ \ldots & \ldots & \ldots & \ldots \\ \Tr ( \Delta Y \{ \rho ,
\Delta \ln \rho - \Delta R \} & ( Y ,  H ) & \ldots & ( Y ,  Y )
\end{array} \rat \, . \label{C.5} \ee

\noindent In order for $ Q $ to vanish, the first column must
vanish, i.e. the operator $ \{ \rho ,  \Delta \ln \rho - \Delta R
\} $ must be the null operator. This condition holds if and only
if $ \rho $ commutes with R, in which case the operator becomes

\bd 2 \sum_i p_i ( \ln p_i - R_i - \sum_k p_k ( \ln p_k - R_k ) )
| i \rangle  \langle  i | \ed

\noindent where $ | i \rangle  $ is an eigenbasis shared by $ \rho
$ and $ R $. It is the null operator if and only if the
eigenvalues $ p_i $ are either $ 0 $ or satisfy equation \ref
{5.21c}:

\bd \ln p_i = R_i - \ln z \ed

\noindent with

\bd - \ln z = \sum_k p_k ( \ln p_k - R_k ) = \sum_k ( 1 / z ) \exp
( R_k) ( R_k - \ln z - R_k ) \, , \ed

\noindent which reduces to equation \ref{5.21d}.

\vspace{\baselineskip}

\break

 \noindent Th4QT

\noindent Equation \ref{5.22} is obtained by substituting equation
\ref{5.16} into the expression for the rate of entropy production

\be d \overline{s} / d t = - k \Tr ( ( d \rho / d t ) \ln \rho )
\label{C.6} \ee

\noindent and making use of the definitions \ref{5.18} and
\ref{5.19}. The important inequality follows from the fact that
the determinant $ g $ defined by equation \ref{5.23} is a Gram
determinant with respect to the scalar product $ (\  , \ ) $, and
it is known that Gram determinants are positive semi-definite (cf.
E.F. Beckenbach and R. Bellman (1965)).

\vspace{\baselineskip} \noindent Th5QT

\noindent This theorem follows directly from the definition of
stable equilibrium state and the fact that such state has the
maximum mean value of the entropy for the given mean values of the
constants of the motion. A mathematical proof of the existence and
uniqueness of such state can be found in A. Katz (1967, pp.
47-50), where, however, the interpretation of the mathematical
symbols is radically different from the present.

\vspace{\baselineskip}

\noindent We conjectured that the statement that we have used to
define stable equilibrium state should follow as a theorem of the
proposed dynamics and the mathematical definition of stability
according to A. Liapunov (1892). In our context, such definition
would read as follows.

\vspace{\baselineskip}

\noindent Definition: Stable Equilibrium States

\begin{quote}
An equilibrium state operator $ \rho' $ is said to represent a
stable equilibrium state if, and only if, for any $ \varepsilon  >
0 $ there exists a $ \delta > 0 $ such that any state operator $
\rho'' $, sharing with $ \rho' $ the values of the constants of
the motion and such that its \emph{distance} from $ \rho' $ is
less than $ \delta $, evolves according to the equation of motion
\ref{5.16} in such a way that its distance from $ \rho' $ at any
later time remains always less than $ \varepsilon $.
\end{quote}

\noindent Here the \emph{distance} between two state operators $
\rho' $ and $ \rho'' $ can be defined as

\be d ( \rho' ,  \rho'' ) = { ( \sum_{i j} { | \rho_{i j}' -
\rho_{i j}'' | }^2 ) }^{1 / 2} \, . \label{C.7} \ee

\noindent However, we have not been able to prove this conjecture.

\vspace{\baselineskip}

\noindent Th6QT

\noindent The mathematical proof of this theorem can be found in
A. Katz (1967, pp. 45-47). It follows from the maximization of the
mean entropy formula subject to the constraint that the state
operator have the given mean values of the constants of the
motion.

\vspace{\baselineskip}

\noindent Th7QT

\noindent A mathematical proof of this theorem can be found in A.
Katz (1967, pp. 50-51).

\vspace{\baselineskip}

\noindent Th8QT

\noindent By substitution of eq. \ref{5.37} into eq. \ref{5.30},
the first term of the equation of motion reduces to the
Schr\"odinger equation for the composite system. The second term
vanishes. To see this, we observe that eq. \ref{5.35} holds for
the operator $ W ( J ) $ and $ \overline{s} ( J ' ) $ vanishes.
Thus, for each $ J $, the determinant of eq. \ref{5.31} vanishes,
the reason being the same as that discussed for theorem Th1QT.

\vspace{\baselineskip}

\noindent Th9QT

\noindent The proof of this theorem is analogous to that of
theorem Th2QT, where instead of the single determinant in eq.
\ref{C.2} there is one determinant for each elementary
constituent.

\vspace{\baselineskip}

\noindent Th10QT

\noindent The proof of this theorem is identical to that of
theorem Th4QT and follows from the known properties of Gram
determinants.

\vspace{\baselineskip}

\noindent Th11QT

\noindent See proof of Th6QT.

\vspace{\baselineskip}

\noindent Th12QT

\noindent See proof of Th7QT.

\vspace{\baselineskip}

\noindent Th13QT

\noindent Equation \ref{5.43} is obtained immediately by partial
tracing equation \ref{5.30} over the complementary factor Hilbert
space of subsystem $ K $, i.e. over all the factor Hilbert spaces
of the elementary constituents that are not contained in subsystem
$ K $. That the terms in the sum of equation \ref{5.30}
corresponding to such elementary constituents vanish under partial
tracing, follows from the fact that the first row of the
determinants in eq. \ref{5.31} vanishes under the trace due to
definition \ref{5.32}.

\vspace{\baselineskip}

\noindent Th14QT

\noindent Consider a system composed of subsystems $ K $ and $ K '
$. Since the two subsystems are separable and in independent
states, equations \ref{5.8}, \ref{5.11} and \ref{5.12} hold. By
substitution into the first term of eq. \ref{5.30} (the
Hamiltonian term), it is seen that such term satisfies eq.
\ref{5.44}. To see that also the second term satisfies eq.
\ref{5.44}, we write it as follows

\be \lbl \begin{array}{c} \sum_{J ( K ) = 1}^{M ( K )} \tau ( J (
K ) ) \{ D ( J ( K ) ) ,  \rho ( J ( K ) ) \} \otimes \rho ( J ( K
) ' ) + \\  \\ \sum_{J ( K ' ) = 1}^{M ( K ' )} \tau ( J ( K ' ) )
\{ D ( J ( K ' ) ) ,  \rho ( J ( K ' ) ) \} \otimes \rho ( J ( K '
) ' ) \end{array} \rbl \label{C.8} \ee

\noindent and observe that, by eq. \ref{5.11},

\be \lbl \begin{array}{ccc} \rho ( J ( K ) ' ) & = & \rho ( J ' (
K ) ) \otimes \rho ( K ' ) \\  &  &  \\ \rho ( J ( K ' ) ' ) & = &
\rho ( J ' ( K ' ) ) \otimes \rho ( K ) \end{array} \rbl \, .
\label{C.9} \ee

\noindent Finally, by eqs. \ref{5.8} and \ref{5.12}, the
determinants $ D ( J ( K ) ) $ (and analogously $ D ( J ( K ' ) )
$), as defined by eqs. \ref{5.31}-\ref{5.34}, are identical to the
determinants that would be associated with the subsystem $ K $
(and $ K ' $) considered as isolated. Thus, substitution of eq.
\ref{C.8} into eq. \ref{C.9} verifies eq. \ref{5.44}.

\vspace{\baselineskip}

\noindent Th15QT

\noindent The if part follows directly from eq. \ref{5.44} and the
evaluation of the rate of entropy production (eq. \ref{C.6}). To
prove the only if part assume that the two subsystems are not
separable. This means that eq. \ref{5.8} is not verified. Thus,
also eq. \ref{5.28} is not verified and the determinants $ D ( J (
K ) ) $ of eq. \ref{C.7} differ from those for subsystem $ K $
considered as isolated. Moreover, the Hamiltonian term of eq.
\ref{5.30} does not split into two terms, since the two subsystems
are interacting. Eq. \ref{5.44} is not verified, therefore, eq.
\ref{5.45} does not hold and the subsystems that were initially in
independent states evolve into dependent states.

\chapter*{REFERENCES} \label{bibl}
\addcontentsline{toc}{chapter}{REFERENCES}

\noindent Band, W. and J.L. Park (1976), \emph{Found. Phys. \bf
6}, 249 (1976).

\vspace{\baselineskip}

\noindent Beckenbach, E.F. and R. Bellman (1965),
\emph{Inequalities}, Springer-Verlag: New York, 1965.

\vspace{\baselineskip}

\noindent Boltzmann, L. (1871), \emph{Wien. Ber. \bf 63}, 397,
679, 712 (1871).

\vspace{\baselineskip}

\noindent Fano, G. (1967), \emph{Mathematical Methods of Quantum
Mechanics}, (English translation) , McGraw-Hill: New York, 1971.

\vspace{\baselineskip}

\noindent Gibbs, J.W. (1902), \emph{Elementary Principles in
Statistical Mechanics}, reprinted in Collected Works and
Commentary, Yale Univ. Press: New Haven, 1936.

\vspace{\baselineskip}

\noindent Glandsorff, P. and I. Prigogine (1971),
\emph{Thermodynamic Theory of Structure, Stability and
Fluctuations}, Wiley-Interscience: New York, 1971.

\vspace{\baselineskip}

\noindent Gyftopoulos, E.P. and G.N. Hatsopoulos (1980), in
\emph{Thermodynamics: Second Law Analysis}, R.A. Gaggioli, ed.,
ACS Symposium Series {\bf 122}, 1980.

\vspace{\baselineskip}

\noindent Hatsopoulos, G.N. and E.P. Gyftopoulos (1976),
\emph{Found. Phys. \bf 6}, 15, 127, 439, 561 (1976).

\vspace{\baselineskip}

\noindent Heisenberg, W. (1927), \emph{Z. Physik \bf 43}, 172
(1927).

\vspace{\baselineskip}

\noindent Jaynes, E.T. (1957), \emph{Phys. Rev. \bf 106}, 620;
{\bf 108}, 171 (1957).

\vspace{\baselineskip}

\noindent Jaynes, E.T. (1978), article in \emph{The Maximum
Entropy Formalism}, R.D. Levine and M.Tribus, eds., MIT Press:
Cambridge, Mass., 1979.

\vspace{\baselineskip}

\noindent Katz, A. (1967), \emph{Principles of Statistical
Mechanics. The Information Theoretic Approach}, W.H. Freeman and
Co.: San Francisco, 1967.

\vspace{\baselineskip}

\noindent Keenan, J.H., G.N. Hatsopoulos and E.P. Gyftopoulos
(1972), \emph{Principles of Thermodynamics}, article in the
Encyclopaedia Britannica: Chicago, 1972.

\vspace{\baselineskip}

\noindent Kossakowski, A. (1972a), \emph{Bull. Acad. Sci. Math.
\bf 20}, 1021 (1972).

\vspace{\baselineskip}

\noindent Kossakowski, A. (1972b), \emph{Rep. Math. Phys. \bf 3},
247 (1972).

\vspace{\baselineskip}

\noindent Kuhn, T.S. (1962), \emph{The Structure of Scientific
Revolutions}, 2nd ed., Chichago Univ. Press: Chicago, 1970.

\vspace{\baselineskip}

\noindent Levine, R.D. and M. Tribus, eds. (1978), \emph{The
Maximum Entropy Formalism}, MIT Press: Cambridge, Mass., 1979.

\vspace{\baselineskip}
 \noindent Liapunov, A.M. (1892), in
\emph{Annals of Mathematics} (French translation), \emph{Studies
\bf 17}, Princeton Univ. Press: Princeton, New Jersey, 1949.

\vspace{\baselineskip}

\noindent Lieb, E.H. and M.B. Ruskai (1973), \emph{Phys. Rev.
Lett. \bf 30}, 434 (1973).

\vspace{\baselineskip}

\noindent Lindblad, G. (1976), \emph{Comm. Math. Phys. \bf 48},
119 (1976).

\vspace{\baselineskip}

\noindent Margenau, H. (1950), \emph{The Nature of Physical
Reality}, McGraw-Hill: New York, 1950.

\vspace{\baselineskip}

\noindent Margenau, H. and J.L. Park (1973), \emph{Found. Phys.
\bf 3}, 19 (1973).

\vspace{\baselineskip}

\noindent Martin, N.F.G. and J.W. England (1981),
\emph{Mathematical Theory of Entropy}, in Encyclopaedia of
Mathematics and Its Applications {\bf 12}, G.C. Rota, ed.,
Addison-Wesley: Reading, Mass., 1981.

\vspace{\baselineskip}

\noindent Maxwell, J.C. (1859), in \emph{The Scientific Papers of
James Clerk Maxwell}, W.P. Niven, ed., Dover: New York, 1965.

\vspace{\baselineskip}

\noindent von Neumann, J. (1932), \emph{Mathematical Foundations
of Quantum Mechanics}, (English translation), Princeton Univ.
Press: Princeton, New Jersey, 1955.

\vspace{\baselineskip}

\noindent Park, J.L. (1968a), \emph{Am. J. Phys. \bf 36}, 211
(1968)

\vspace{\baselineskip}

\noindent Park, J.L. (1968b), \emph{Phil. Sci. \bf 35}, 205, 389
(1968).

\vspace{\baselineskip}

\noindent Park, J.L. and W. Band (1977a), \emph{Found. Phys. \bf
7}, 233 (1977).

\vspace{\baselineskip}

\noindent Park, J.L. and W. Band (1977b), \emph{Found. Phys. \bf
7}, 813 (1977).

\vspace{\baselineskip}

\noindent Park, J.L. and W. Band (1978), \emph{Found. Phys. \bf
8}, 239 (1978).

\vspace{\baselineskip}

\noindent Schr\"odinger, E. (1936), \emph{Proc. Camb. Phil. Soc.
\bf 32}, 446 (1936).

\vspace{\baselineskip}

\noindent Shannon, C.E. (1948), \emph{Bell System Tech. J. \bf 27}
(1948); cf. also C.E. Shannon and W. Weaver, \emph{Mathematical
Theory of Communication}, University of Illinois Press: Urbana,
1949.

\vspace{\baselineskip}

\noindent Simmons, R.S. (1979), Ph.D. Thesis, Washington State
University, 1979.

\vspace{\baselineskip}

\noindent Strocchi, F. (1966), \emph{Rev. Mod. Phys. \bf 38}, 36
(1966).

\end{document}